\documentclass{article}

\usepackage{microtype}
\usepackage{graphicx}
\usepackage{subcaption}
\usepackage{booktabs} % for professional tables
\usepackage{caption}
\usepackage[subrefformat=parens,labelformat=parens]{subcaption}
\usepackage{tikz, amsmath, filecontents, algorithm, amsmath, epsfig,endnotes,enumitem,makecell,multirow, microtype, xspace, setspace, authblk, xcolor,textcomp,graphicx,filecontents,caption,subcaption, url,tabularx, times, subcaption,float, enumitem, listings
}

\usepackage[dvipsnames]{xcolor}

\newcommand{\sysname}{Vortex\xspace}
\newcommand{\rdmalib}{Flash\xspace}

\usepackage{etoolbox}

\usepackage[accepted]{mlsys2025}

\AtBeginDocument{%
  \setlength{\textfloatsep}{4pt plus 1pt minus 1pt}%
  \setlength{\floatsep}{4pt plus 1pt minus 1pt}%
  \setlength{\intextsep}{4pt plus 1pt minus 1pt}%

  \captionsetup[table]{skip=2pt, belowskip=0pt}

  \captionsetup[subfigure]{aboveskip=1pt, belowskip=1pt}
   \captionsetup[figure]{skip=-8pt, belowskip=-8pt}
  \setlist[itemize]{noitemsep, topsep=0pt, partopsep=0pt, parsep=0pt, leftmargin=1.5em}
}
\mlsystitlerunning{Submission and Formatting Instructions for MLSys 2025}

\begin{document}

% \textsuperscript{1,*}, \textsuperscript{1,*}, \textsuperscript{1}, \textsuperscript{2},\\
% Timothy Louie\textsuperscript{1}, \textsuperscript{1}, \textsuperscript{1}, \textsuperscript{1}\\
% \vspace{0.5em}
% \textsuperscript{1,$\dagger$}, \textsuperscript{1}

% \\
% \textsuperscript{1}Cornell University, \{yy354, ty373, jah649, tjl96, jq54, az275, yw2399, ws393, ken\}@cornell.edu\\
% \textsuperscript{2}University of Oslo, thiagoga@ifi.uio.no

\twocolumn[
\mlsystitle{{\sysname}: Hosting ML Inference and Knowledge Retrieval Services  With Tight Latency and Throughput Requirements}
\mlsyssetsymbol{equal}{*}
\begin{mlsysauthorlist}
\mlsysauthor{Yuting Yang}{equal,cornell}
\mlsysauthor{Tiancheng Yuan}{equal,cornell}
\mlsysauthor{Jamal Hashim}{cornell}
\mlsysauthor{Thiago Garrett}{oslo}\\
\mlsysauthor{Jeffrey Qian}{cornell}
\mlsysauthor{Ann Zhang}{cornell}
\mlsysauthor{Yifan Wang}{cornell} 
\mlsysauthor{Weijia Song\protect\textsuperscript{\dag}}{cornell}
\mlsysauthor{Ken Birman}{cornell}
\end{mlsysauthorlist}
\mlsysaffiliation{cornell}{
Cornell University 
\texttt{\{yy354, ty373, jah649, tjl96, jq54, az275, yw2399, ws393, ken\}@cornell.edu}
}
\mlsysaffiliation{oslo}{
University of Oslo 
\texttt{thiagoga@ifi.uio.no}
}
\mlsyskeywords{Machine Learning, MLSys}
\begin{abstract}
There is growing interest in deploying ML inference and knowledge retrieval as services that could support both interactive queries by end users and more demanding request flows that arise from AIs integrated into a end-user applications and deployed as agents.  Our central premise is that these latter cases will bring service level latency objectives (SLOs).  Existing ML serving platforms use batching to optimize for high throughput, exposing them to unpredictable tail latencies.  {\sysname} enables an SLO-first approach.  For identical tasks, {\sysname}'s pipelines achieve significantly lower and more stable latencies than TorchServe and Ray Serve over a wide range of workloads, often enabling a given SLO target at more than twice the request rate.  When RDMA is available, the {\sysname} advantage is even more significant.
\end{abstract} 
]

% this must go after the closing bracket ] following \twocolumn[ ...

% This command actually creates the footnote in the first column
% listing the affiliations and the copyright notice.
% The command takes one argument, which is text to display at the start of the footnote.
% The \mlsysEqualContribution command is standard text for equal contribution.
% Remove it (just {}) if you do not need this facility.

%\printAffiliationsAndNotice{}  % leave blank if no need to mention equal contribution
\printAffiliationsAndNotice{\mlsysEqualContribution} % otherwise use the standard text.
\renewcommand{\thefootnote}{\fnsymbol{footnote}}
\footnotetext[2]{Work done while at Cornell University.}
\renewcommand{\thefootnote}{\arabic{footnote}}

\section{Introduction}

AI-enhanced interfaces and applications are being widely adopted, leading to a growing interest in ML used as a service~\cite{agentsurvey2024feifeili,MArk,MLaaS2022}.  Loads on such services will be higher than in settings where every query originates with a a human user, but latency will matter too (think of AI in equity trading, or playing a role in medical devices).  We anticipate growing demand for SLOs: latency targets with miss-rate limits, a trend that runs counter to a tradeoff that historically prioritized throughput~\cite{zhao2024blendserve,ali2020batch}.  

On the other hand, once an SLO is achieved there is no benefit to driving latency even lower.  Thus we can design ML-as-a-service frameworks that facilitate SLOs while still employing batching to enhance throughput.  The challenge is to avoid queuing backlogs, so that we can sustain high throughput without triggering SLO misses. 

Our work views ML services as pipelines that will run on elastically-resizable pools of servers.  Such a pipeline is simply a directed workflow graph with an ingress and an egress node~\cite{inferline,shepherd,INFaaS,NEURIPS2023_judingllm_mtbenchmark,packer2024memgptllmsoperatingsystems}.  Nodes represent ML stages and directed edges represent data flows. Pipelines can share components: even if the request flow rates are bursty and unpredictable, a shared ML service will often see steadier loads that can be opportunistically aggregated.  

The resulting batches will vary in size and may be smaller than what a throughput-first design would employ, although still large enough to enable efficient use of GPUs.  If different pipelines require some of the same components the corresponding pool of instances can be managed as an elastically resizable pool, much like with web services~\cite{cloudMicroserviceArch, deathstarbenchmark, autothrottle, nightcore, meta_microservice_topology,muCache}.  

\begin{figure}[h]
     \centering
     \begin{subfigure}[b]{0.42\textwidth}
         \centering
        \includegraphics[width=\textwidth]{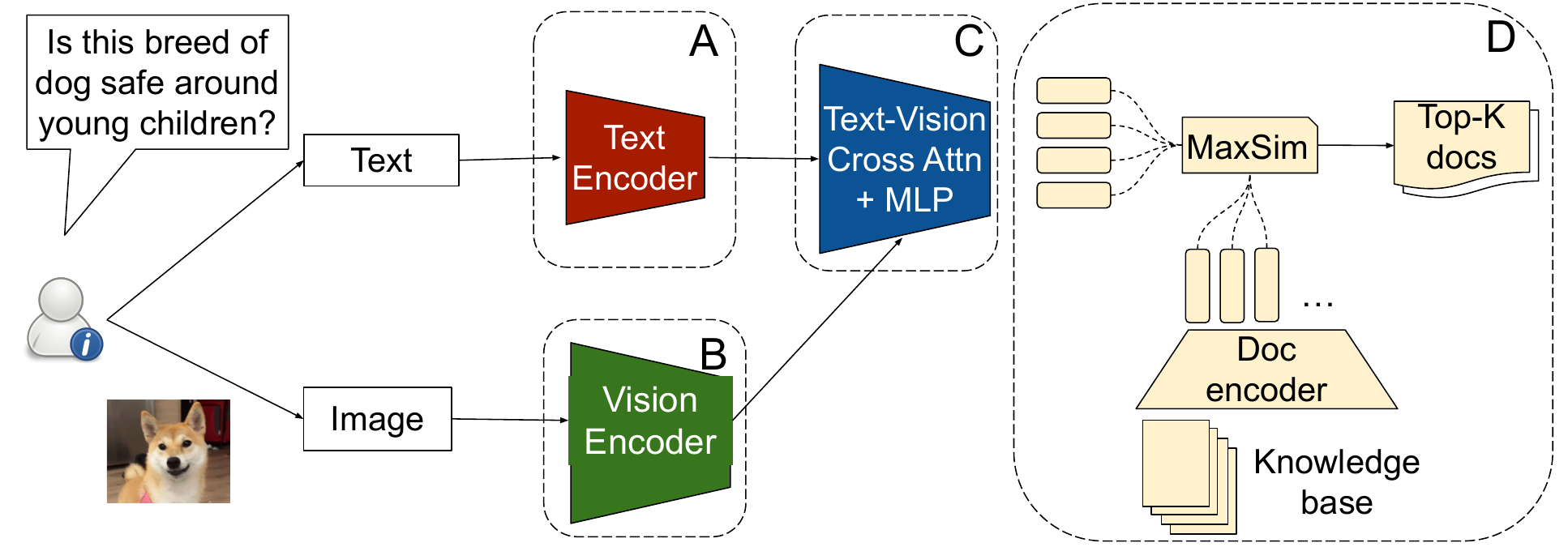}
         \caption{PreFLMR knowledge retrieval pipeline.}
         \label{fig:pipeline1_flmr}
     \end{subfigure}
     \hfill
     \begin{subfigure}[b]{0.45\textwidth}
         \centering
         \includegraphics[width=\textwidth]{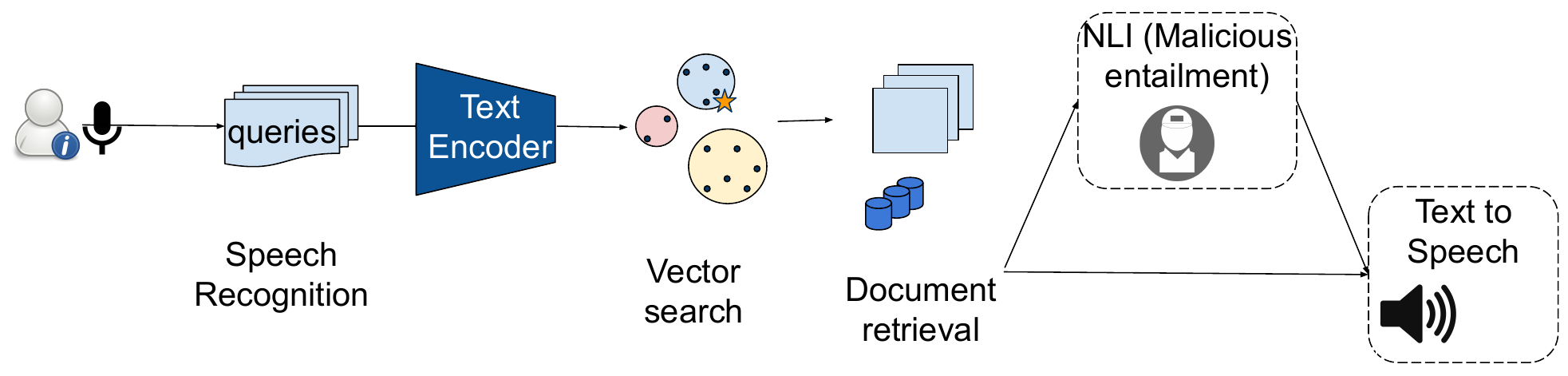}
         \caption{AudioQuery knowledge retrieval pipeline.}
         \label{fig:pipeline2_speechRAG}
     \end{subfigure}
\vspace{-1.0em} 
\caption{Two Representative ML-as-a-Service Pipelines}
\label{fig:pipelines}
\end{figure}

We use two ML pipelines as running examples (Figure~\ref{fig:pipelines}): 

(1) {\bf PreFLMR}~\cite{lin2024preflmr} is a knowledge retrieval application that takes an image and an associated query and retrieves pertinent documents.  

(2) {\bf AudioQuery} is a speech-query RAG LLM\footnote{We refer to transformer-based models like BERT~\cite{devlin2019bert}, BART~\cite{lewis2019bart} and RoBERTa~\cite{liu2019roberta} as Large Language Models throughout the paper even when they have no generative role} pipeline that converts an audio query to text, searches for documents relevant to the query, and then generates a spoken response after filtering for undesired language. \\

% \newpage
Our contributions are as follows:

\textbf{Microservice‑based pipeline architecture.}  
    {\sysname} employs a novel DLL-based extension architecture.  ML pipelines can be hosted in our address space, with each ML component treated as a trusted tenant.  When an ML accesses 
    an input or data object pointers are used, minimizing copying and network transfers.
    
    {\sysname} servers play double duty: each  can be configured as both a key-value storage server and as a compute host.  By having our scheduler route queries to components running where the data they depend upon already resides (such as models, vector database indices, etc), we minimize access delays. This goal aligns nicely with {\em sharding:} {\sysname}'s key-value storage space is split into groups of replicas for scalability and fault-tolerance. By aligning component pools with shards, we can vary the component pool size up to the full set of shard members without copying dependencies over the network.  An {\em affinity grouping} feature ensures that objects accessed as a set are collocated on the same shard and jointly loaded or evicted from cache.

\textbf{We identify other sources of latency spikes.}  
    For SLO-oriented systems, queuing delays and excessively large batch sizes are problematic, yet batching remains central to high throughput.  {\sysname} batches opportunistically.  Our elasticity mechanism preloads models and other dependent objects into GPU memory before 
    activating new instances, avoiding another significant source of delay.

\textbf{System‑level optimizations.}  
    We identify additional optimizations that benefit SLOs:  {\bf (1) Smart task placement:} In a pipelined architecture, we have freedom to place components on the same or different machines,  Our scheduler seeks to minimize delays by collocating components on the same machine if the predicted stage-to-stay delay would be high and the machine has adequate capacity to run both side by side. (2) {\bf Pool-oriented microservice management}: We use anticipated workload and microbenchmarking to ``right-size'' each pool for its particular load and compute requirements, and to limit opportunistic batches to SLO-compatible sizes. {\bf (3)  Zero-copy data paths.} We develop an asynchronous  architecture that avoids copying and minimizing locking on its end-to-end data paths, encouraging continuous flow of requests and data: a technique that proves beneficial both on TCP networks and on RDMA.   {\bf (4) Smart packing.}  We identify unusually efficient strategies for mapping components to GPU, with striking gains relative to monolithic pipeline deployments.

Jointly, these techniques  enable {\sysname}-hosted ML services to offer tight SLOs with high throughput.  Using the same SLO, our throughput is often double that of Ray Serve (both systems far outperform Torch Serve).
    
%\textbf{Improved performance.}  
%We benchmark PreFLMR and AudioQuery on TorchServe, Ray Serve and {\sysname}. {\sysname} significantly improves latency while matching the throughput of Ray Serve, one of today's most widely used platforms (both are far faster than TorchServe).   %Ann: I'm not sure how much detail we want to include here (probably not much), but I thought the story was more about tail latency than throughput? Do we want to throw that in?

 %Ken: I feel as if these next lines duplicate both the abstract and the last of the contributions.  Be wary of repetition!  Reviewers do notice it.
 % {\sysname} includes a number of system-level optimizations: compute collocation, efficient stage-to-stage handoffs, data caching, and adaptive opportunistic batching, all intended to mitigate overheads and improve end-to-end efficiency.   Our evaluation explores the exact value of each, individually and in combination.   We show that the {\sysname} microservices approach achieved 20-40\% higher throughput than any of the monolithic deployments (on equivalent hardware), and also that {\sysname} achieves higher throughput and lower latency (with sharply lower tail latency) than Ray Serve in its streaming configuration.  Interestingly, we also find that when load is light, {\sysname} achieves  latencies under light load that are closer to monolithic deployments, whereas the componentized deployments of the comparison systems perform considerably worse than their monolithic options for this scenario. 

\section{Background}
\label{sec:background}
ML inference systems generally embed text, images and other query data to obtain high dimensional vector representations, then perform inference using computational kernels built on linear algebraic primitives (primarily, matrix multiplication).  The solution might entail ANN queries to retrieve relevant documents, running code, or launching agents.   %In pipelines like the ones used by PreFLMR~\cite{lin2024preflmr} and AudioQuery,  objects passed from stage to stage would be embedding vectors, perhaps batched into matrices in which each row is an embedding.  To a lesser degree, pipelines also  pass files holding documents or images, image markups, etc.  Additionally, each ML stage depends on one (or more) pretrained ML models, and RAG LLMs depend on preindexed databases of documents.  Rather than passing these dependent objects around, they would often be loaded from storage. 

Against this overall backdrop, recent developments set the stage for our work, starting with a shift from monolithic AI designs to componentized data flows, which can be deployed monolithically or as pipelines of interacting components.  The shift is occurring primarily because foundation models are so costly to train that developers of new MLs often prefer to start with general purpose off the shelf components and link them into new pipelines, which can then be fine-tuned.  There is also growing interest in deploying MLs as a service, motivating the interest in SLOs. 

A componentized approach potentially brings costs: such MLs potentially have many points at which copying might occur when data is passed from one component to another.  RAG MLs, MLs that execute code, and MLs that depend on object other than the nodel all have steps at which a stall can arise, threatening SLO misses in the pipeline as a whole.  Moreover, when a queuing delay arises, it can ripple through the entire pipeline.  Yet componentization also brings opportunities that {\sysname} explores.

Today's serving systems often have highly asynchronous designs.  Such designs are prone to queuing backlogs, but here one finds unexpected help from mathematics: the GEMM  matrix methods prevalent in ML scale sub-linearly in the batch size of the input query matrix.  Accordingly, most MLs batch, processing sets of concurrent queries by concatenating the query embeddings as rows in a query matrix.
Our work arises in the same context, but rather than maximizing batch sizes whenever possible {\sysname} actively manages backlogs and limits batch sizes with the goal of avoiding SLO misses, elastically resizing ML microservices if extra capacity is needed.   

\section{Representative ML Pipelines}
The two MLs we use as running examples are typical multimodal ML inference and knowledge retrieval pipelines.
\subsection{An Existing Componentized Multimodal ML}
\label{sec:pflmr}
FLMR~\cite{pmlr-v139-kim21k} is a popular application for answering questions about images.  Although normally used as a monolithic program it is internally componentized.  A pipeline called PreFLMR~\cite{lin2024preflmr} is used to retrieve relevant documents, which FLMR feeds to its response generator. We isolated  PreFLMR for deployment as a document retrieval service.  Given (query, image) pairs it returns identifiers for the most relevant documents.  The pipeline starts by encoding the inputs: text using an LLM transformer and images using a visual encoder.  The visual stage applies a ``masked language'' operation to focus attention on relevant aspects of the segmented image, resulting in a query to a Colbert search for matching documents~\cite{colbert_original, colbert_serve}.  
\subsection{A Hand-Built Pipeline}
We created AudioQuery as a knowledge-retrieval tool for clients who will issue vocal queries related to document corpi, news feed, etc.  Inputs are spoken and responses are vocalized. As seen in  (Figure~\ref{fig:pipeline2_speechRAG}) it uses several pretrained general-purpose foundation models: An audio-to-text model~\cite{an2024funaudiollm}, the BGE model~\cite{chen2024bge} to embed the query, and then a vector search that employs a FAISS index to retrieve relevant documents~\cite{faiss}.  To filter the outputs, we first
classify emotional tone using BART~\cite{lewis2019bart} fine-tuned on GoEmotions\cite{demszky2020goemotions} (sourced from Reddit and labeled for 27 emotions). The last stage is a Text-to-Speech (TTS) component based on NVIDIA’s FastPitch model~\cite{lancucki2021fastpitch}.

\section{Platform Design}
\label{sec:platform_design}
\subsection{{\sysname} Architecture}

\begin{figure}[ht]
    \centering
    \includegraphics[width=.45\textwidth]{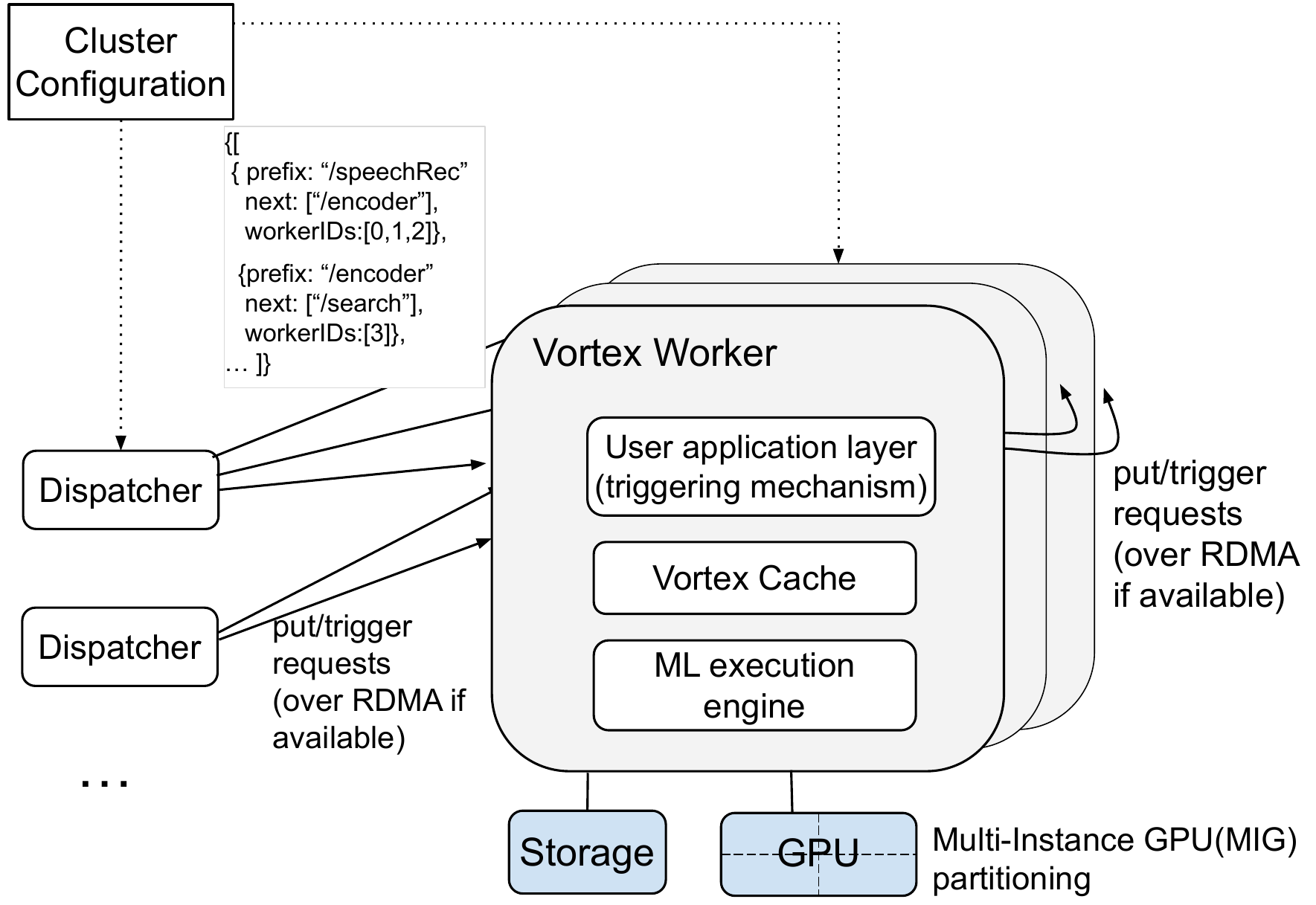}
    \vspace{-1.0em} 
    \caption{{\sysname} System Architecture}
    \label{fig:arch}
\end{figure}

{\sysname}'s overall architecture is illustrated in Figure~\ref{fig:arch}.  

{\sysname} servers perform two simultaneous roles: they host a key-value distributed data store (KVS) and they host and execute ML logic. The KVS offers the usual {\bf put} and {\bf get} API, with control over object persistence and supporting time-indexed data retrieval. Optional wrappers offer standard POSIX file system APIs and the Kafka DDS and queuing middleware API, mapping both to our KVS framework so that when a hosted ML interacts with external data, data paths route through our framework.   

%The {\sysname} developer uses a configuration file to express the mapping of this KVS to hardware, which object pools reside on which machines, and desired properties for each, including the replication factor.  For example, with a factor of 3 each shard would be replicated on 3 servers (their data will be identical, but cache contents can differ).  

In its ML hosting role, {\sysname}'s central goal is to route computational tasks to server instances that already have the needed ML models and data (indeed, that have already loaded them into GPU memory). If an application accesses non-local data {\sysname} will fetch it, then retain it in cache.% (the retention policies and algorithms for cache-aware approach task placement are detailed in~\cite{Navigator}). 

At startup each {\sysname} server loads a list of ML components, packaged as dynamically linked libraries (DLLs). Each holds ML logic and a copy of any ML package it requires.  ML models and other required data objects are saved into our KV store, leveraging an {\em affinity grouping} mechanism~\cite{affinity} that collocates objects that will be needed simultaneously.   

When first loaded, a DLL registers one or more {\em triggers}.  Each specifies a key or a pathname prefix that the DLL wishes to monitor and a callable function: if a {\bf put} occurs on a watched key (creating a new object, or a new version of an existing object), an upcall from {\sysname} to the function will occur on every replica (in the identical order), passing pointers to the key and the value as arguments.  Often we wish to trigger ML compute without storing an object, so the KVS API also includes a {\bf trigger} put operation.  This comes in two variants.  A {\em routed} {\bf trigger} allows the caller to designate a specific server.  A {\em load-balanced } {\bf trigger} automatically randomizes over the shard members. 

\subsection{Network Accelerators}
RDMA networking is widely used during ML training, yet uncommon for ML serving:  operators see it as costly and fragile, and limit its use to high-performance compute (HPC) clusters.  Despite this, our work reflects the belief that as ML becomes a service, RDMA or RoCE will become important because it can drive SLOs down beyond anything achievable in software. 
Accordingly, we built  {\sysname} over {\rdmalib}[anonymized], a layer that leverages RDMA when possible (using TCP if not), and offers a variety of zero-copy communication primitives for point-to-point, multicast, atomic multicast and durable atomic multicast (Paxos).  It turns out that the design choices required for RDMA yield benefits in {\sysname} even we run purely on TCP (Section~\ref{sec:experiments}).

The {\rdmalib} layer of {\sysname} guarantees fault-tolerance (by reissuing queries or updates disrupted by a failure in a manner that guarantees exactly-once semantics) and supports a formal consistency model: {\em serializable snapshot isolation}~\cite{SerializableSnapshotIsolation}: Stages of an application read what prior stages wrote, preserving update ordering.  Under this model, replicated data is updated atomically and reads are guaranteed to see the most current data.   Today strong consistency guarantees are rarely needed in ML systems, but future MLs will run in settings where real-time updates will be more prevalent.  Stronger assurances may then grow in importance.  The topic is explored in Appendix A.

\section{Challenges and Opportunities}
\label{sec:challenges}
Broadly, we now have the main elements of our approach: the developer creates or obtains an ML pipeline, connecting stages using {\bf trigger put} and {\bf put} operations (or with POSIX file sharing, or Kafka), configures {\sysname} to load the DLLs, uploads required ML models, document and vector database indices into the KVS, and then clients can query the service.  

\subsection{SLO Targets}
\label{sec:SLO_targets}
Although other papers have explored SLO-oriented task scheduling and auto-throttling~\cite{wang2024autothrottle,romero2021infaas,crankshaw2017clipper,sinan-asplos2021,dean2013tail}, our ground-up SLO-first approach is novel.  At the same time, throughput remains important: batching makes services more cost-effective.  Thus our expectation is that platform owners will will seek the most cost-effective service offering that can still satisfy their latency targets and tolerance of SLO misses, and will identify cost-effectiveness with throughput.  The use case will often determine how the SLO/throughput tradeoff should be made.  Physical-control applications (such as in robotic surgical devices) are likely to treat latency overruns as major faults.  Human-user interactive applications might see SLO misses as merely inconvenients.  

To accomplish this goal, we enable a methodology in which the ML service can advertise a {\em model} of how its latency and miss rates behave as a function of load.  The user can then select a desired operating point, which is used to configure the {\sysname} scheduler to elastically resize component pools, keeping loads within the target range within which ML service-level SLOs can be satisfied.

%Additionally some steps have large output data size to be send to the next stage. 
\subsubsection{Opportunistic batching}

\begin{figure}[ht]
 \centering
  \includegraphics[width=0.45\textwidth]{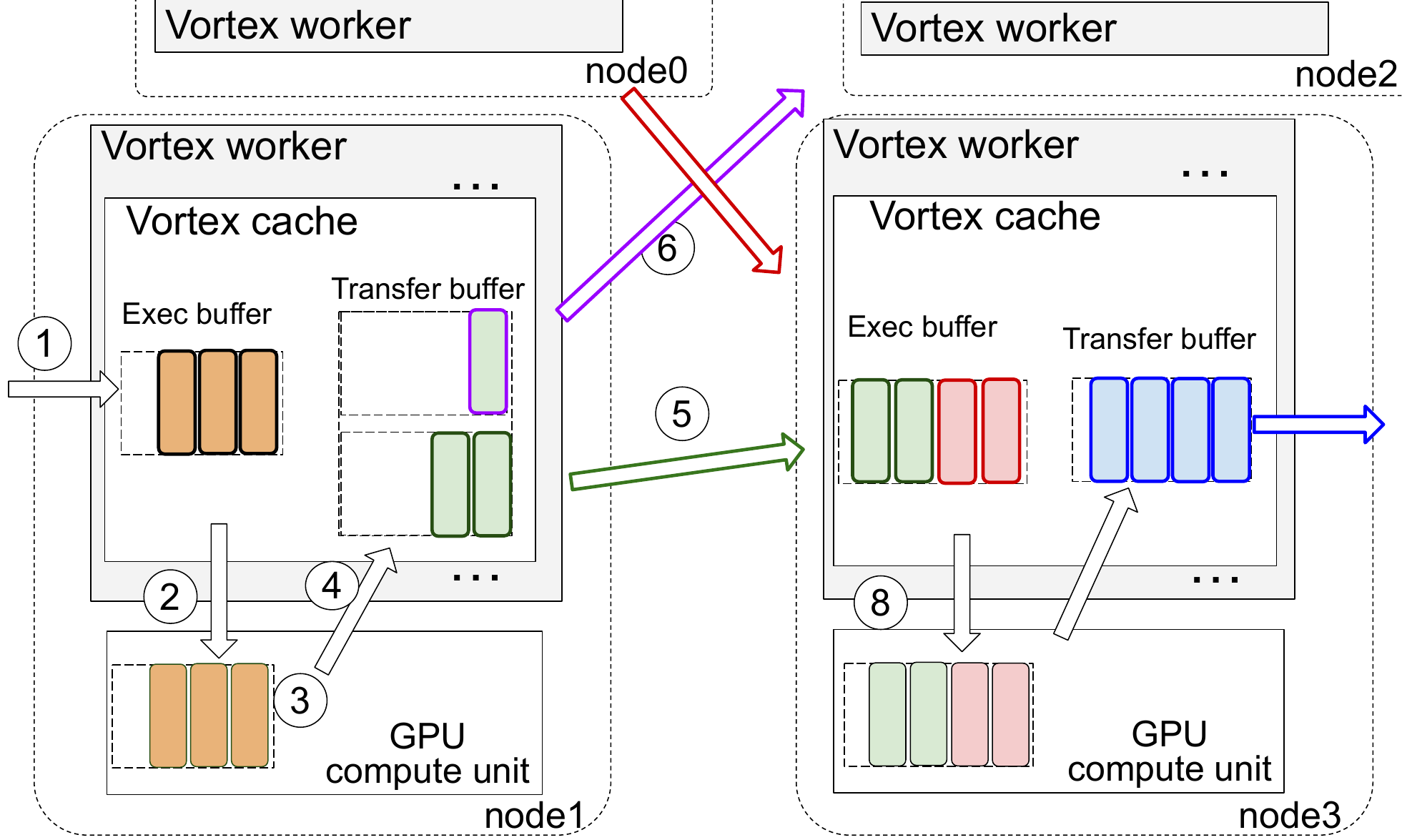}
 \vspace{-1.0em} 
 \caption{Stage to Stage Handoffs}
 \label{fig:dynamic_batch}
\end{figure} 
%We noted that in a monolithic pipeline, opportunistic batching occurs each time a new task launches: the ingress node scoops up the currently waiting queries, up to some maximal batch size, but that this results in significant inefficiencies.  In contrast, a multi-stage deployment can employ opportunistic batching stage by stage, which proves to be more effective because each stage typically has a different optimal batch size  (as was shown in Section~\ref{sec:diff_resource_requirements}), and because delays and backlog can easily arise within the pipeline: with a fixed batch the pipeline lacks flexibility, but  opportunistic batching at a stage-to-stage level can often catch up. %For example, step X may see throughput advantages until a batch size of 4, and step Y until a size of 32. In a monolithic deployment focused on throughput, step X would be forced to run in batch sizes of 32 even though it increases the latency without helping the throughput for that individual step. Memory and speed can also play roles in the optimal batch size for individual steps – For example, a step that runs substantially faster than others but has linear memory scaling as batch size increases would benefit from smaller batch sizes. A microservice deployment has the flexibility to allow each stage to individually opportunistically batch, allowing for different steps to batch only as much as they need. The difference in observed batch sizes for monolithic vs microservice deployments can be seen in section \ref{sec:exp_batching}  
A first question relates to queuing effects that arise in pipelines.  There are several stages where these are seen, illustrated in  Figure~\ref{fig:dynamic_batch}, and we can apply opportunistic batching in each case: (1) At each stage {\sysname} enqueues queries on a queue of pending work, hence (2) the execution dispatcher for  can remove multiple pending tasks and perform them as a batch (the figure shows a GPU, but the pattern would be the same for host compute).  (3) After execution completes, the results are released to the next stage.  Note that because a component can be shared by multiple pipelines, the results produced by a single component may need to be sent to different next stages.  Additionally, due to elastic pool resizing, the number of currently active workers in each pool could vary, hence a selection of the specific worker to which the output will be sent must occur, offering a scheduling opportunity that we explore below.  (4) The send queue management thread forms a batch of outgoing results and (5) asynchronously sends them to the next stage. (6) Finally, notice that one stage could require inputs from multiple upstream stages, as seen in stage C (cross-attention) in the PreLMR pipeline.  Here we must form {\em matched sets} of inputs corresponding to the same query id prior to passing the batch of work to the GPU.

\subsection{Batch size tuning}
\begin{figure}[ht]
    \centering
    % Row 1 — Pipeline1 (4 plots)
    %\begin{subfigure}[b]{0.45\textwidth}
        %\centering
        % \includegraphics[width=.45\textwidth]{plots/micro_benchmark/nograd_ppl1_throughput_latency_gpu_mem_batch_plot.pdf}
        \includegraphics[width=.48\textwidth]{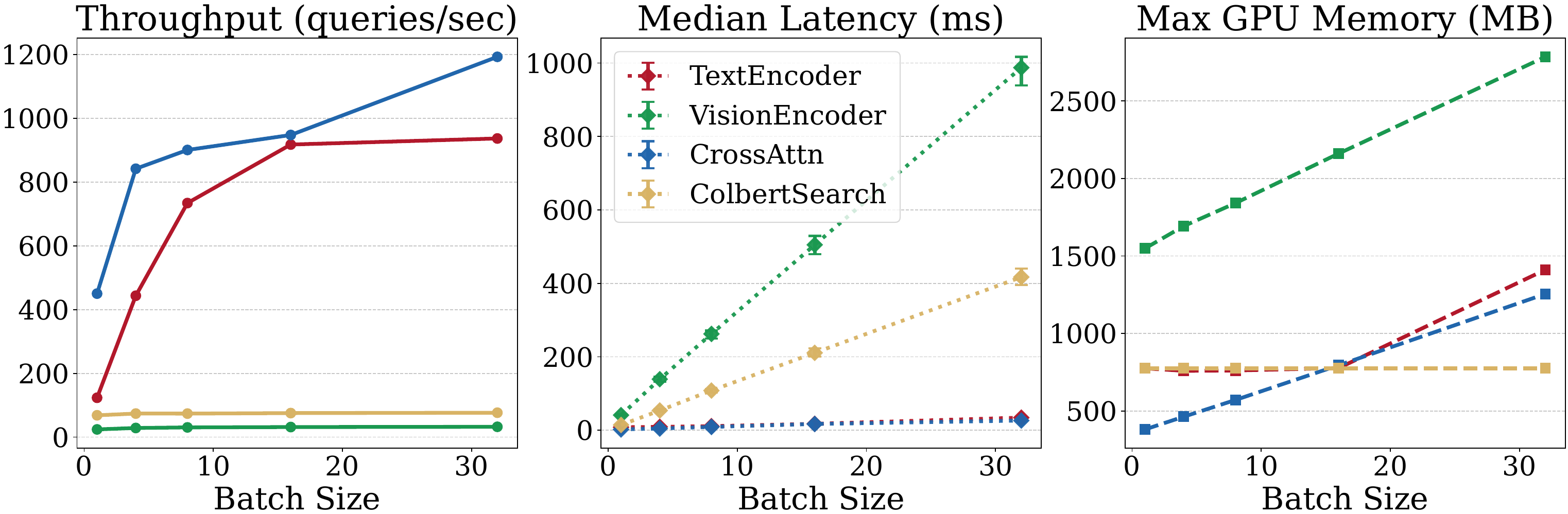}
        % \caption{\centering Requirements for PreFLMR as a function of batch size}
        %\label{fig:pipeline1_throughputs}
    %\end{subfigure}
   
    \vspace{-1.0em} 
    \caption{Resource requirements of PreFLMR components. We vary the batch size and show throughput and GPU memory usage. }
    \label{fig:components_profile}
\end{figure}

\label{sec:diff_resource_requirements}
In the introduction, we noted that if a system doing opportunistic batching forms excessively large batches, some requests might experience SLO violations.  To understand this risk, Figure~\ref{fig:components_profile} shows an experiment in which we create request batches of varying size and measure throughput, latency and GPU memory usage on a stage by stage basis in PreFLMR.  If our only consideration was throughput, we could read off the optimal batch size for each component from these graphs.  Broadly, as batch size increases, performance rises, limited by the available host memory, GPU memory and compute resources.  On the other hand, latency rises because with larger input objects the computation takes longer, and this can penalize a long-waiting query.

Notice that components often reach a peak of efficiency after which larger batches require more memory and yet throughput ceases to improve.  Thus even if latency were not an objective, the sweet spot is when  the pipeline as a whole runs with optimal per-component batch sizes.  

Further, notice that as batches become larger, latency rises.  The point at which SLO violations might occur is not easy to  predict from Figure~\ref{fig:components_profile} because stage-to-stage handoffs have an impact too: a full end-to-end evaluation of the kind we do in Section~\ref{sec:experiments} becomes necessary.  But the implication is that with SLO goals, elasticity is sometimes needed to increase capacity and prevent large backlogs.

\begin{figure}[H]
    % \vspace{1em} 
     \centering
     \begin{subfigure}[b]{\columnwidth}
         \centering
         \includegraphics[width=0.7\columnwidth]{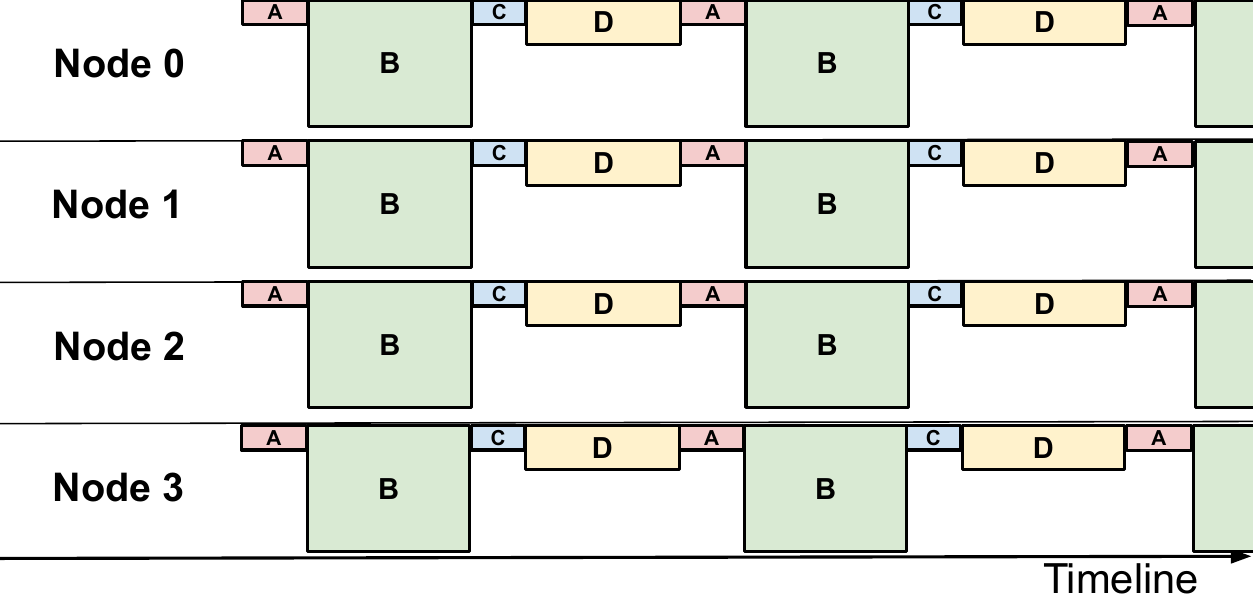}
         \caption{Timeline for Monolithic Deployment of PreFLMR.}
         \label{fig:mono_illustration}
     \end{subfigure}
     % \vspace{1em}
     \begin{subfigure}[b]{\columnwidth}
         \centering
         \includegraphics[width=0.7\columnwidth]{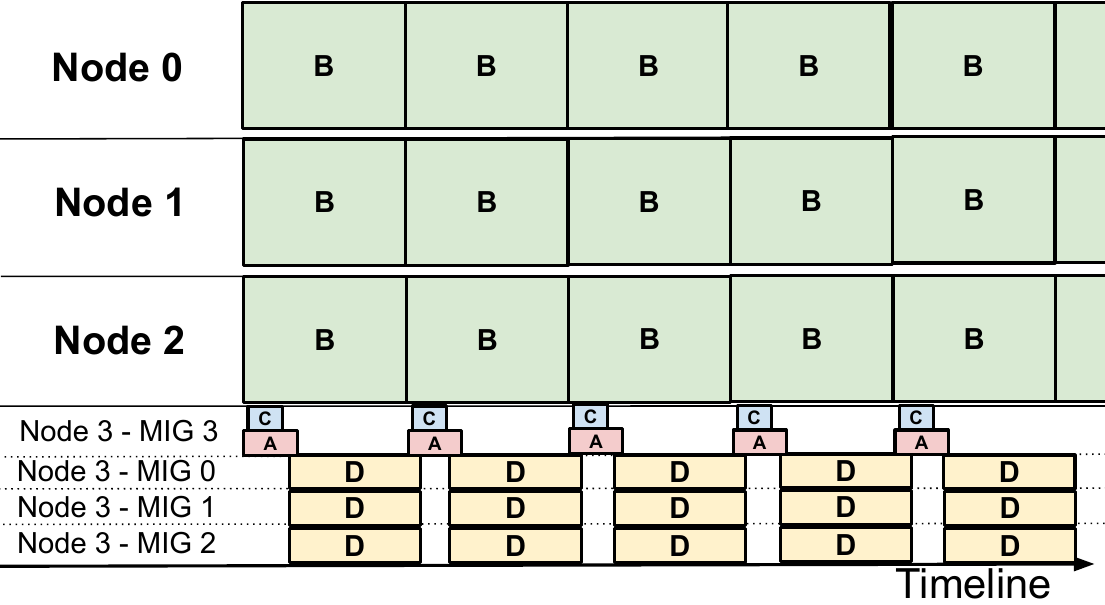}
         \caption{Timeline for Microservice Deployment of PreFLMR.}
         \label{fig:micro_illustration}
     \end{subfigure}
\vspace{-2.2em} 
\caption{Resource packing comparison for two PreFLMR deployment options. With monolithic deployments (top), PreFLMR performs 8 queries in the time period shown.  The microservice option (bottom) enables scheduling flexibility: by running visual embedding (B) on three nodes and the remaining components (A,C,D) on the fourth, throughput rises to 15 queries in the same time period.}
\label{fig:mono_vs_micro_timeslines}
\end{figure}

 \subsection{Elasticity Challenges}
\label{sec:elasticity}
Dynamic resizing for MLs deployed as services pose challenges not seen with traditional web service infrastructures.  In standard cloud settings it is possible to ``spin on a dime'' because launching new microservice instances is inexpensive.  But when a new ML server instance is launched we must load the needed ML model and other required objects into the GPU, from local storage (if the objects are all in the shard where the instance runs) or over the network. Best is to {\em preload} models in anticipation of need, and hence speculatively: early in a load surge we should prepare the new worker node.  Then, if the surge continues we can resize without triggering a model-load stall.  The effect is assessed experimentally in Section~\ref{sec:cold_start}.

Elasticity also raises a somewhat subtle issue related to ordering.  Consider incast patterns of the kinds seen in PreFLMR, where the pipeline concurrently embeds the query and image, but then needs to run Text-Vision Cross Attention on the pair of outputs.  Call the subtasks A, B and C.  As A and B are completed, how can the server know which cross-attention server instance will run C?   Clearly, A and B must reach identical decisions, but this creates a tension relative to the desire for load-balancing.  Our solution is to make load-balancing decisions in the ingress node when requests first reach the pipeline.  We then can tag each incoming request with a unique request id and with our load-balancing choices, effectively locking down the routing A and B will later use.  The approach will also preserve request ordering within a stream of requests.

\begin{figure}[ht]
    \vspace{0.5em} 
    \centering
    \begin{subfigure}[b]{0.38\textwidth}
        \centering
        \includegraphics[width=\textwidth]{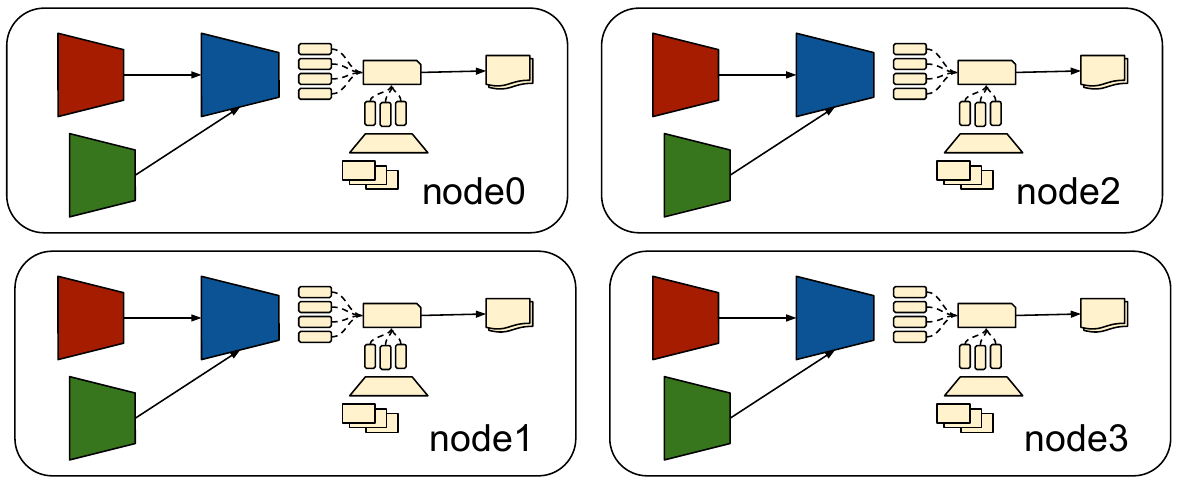}
        \caption{Monolithic Deployment}
        \label{fig:mono_deployment}
    \end{subfigure}%
    \vspace{.3em} 
    \begin{subfigure}[b]{0.38\textwidth}
        \centering
        \includegraphics[width=\textwidth]{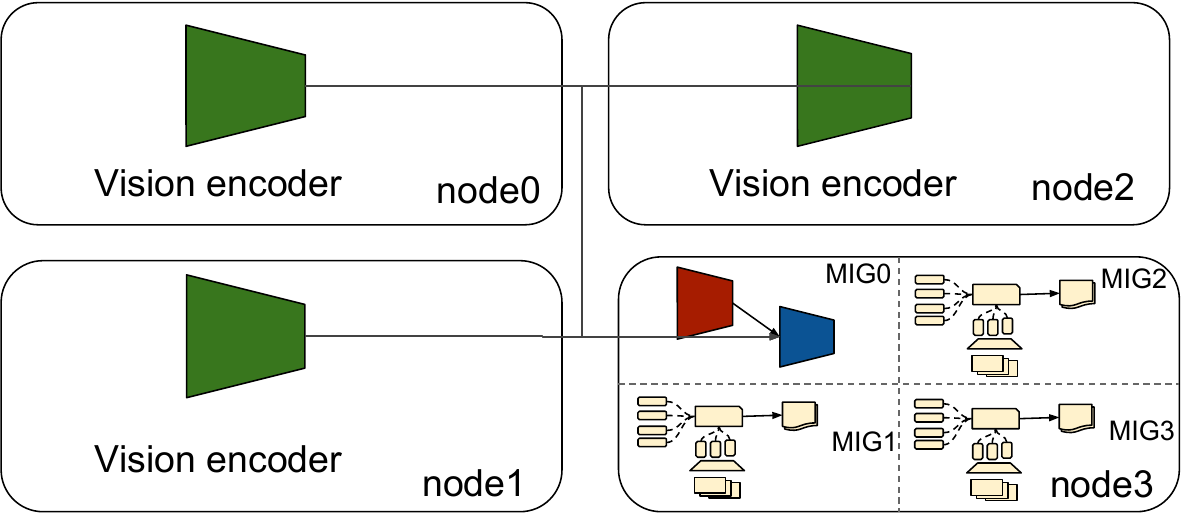}
        \caption{Microservice Deployment}
        \label{fig:micro_deployment}
    \end{subfigure}
    \vspace{-1.0em} 
    \caption{Monolithic vs. Microservice deployments corresponding to the timelines in Figure~\ref{fig:mono_vs_micro_timeslines}.   The top 
    layout is monolithic: all components of PreFLMR are placed on all servers in use. At whatever batch size is
    used, we can run 8 query batches in the time period shown.  The microservice layout (bottom) is more flexible, enabling substantially higher throughput.}
    \label{fig:mono_vs_micro_config}
\end{figure}

%\section{Component Packing}
%\label{sec:methodologies}
\subsection{Packing Microservices on Servers}
\label{fig:deploy_illustration}

% Structure of each subsection 
% 1. intuition/observation that explain why do we need that methodology (why do we need that technique)
% 2. How the techniques can help (what are lacking in the other ways)
% 3. technical explanation of how to achieve that methodology

We next consider optimizations that collocate  multiple components on the same servers.  Ineffective packings leave resources idle because of task sequencing in the pipelines.  The issue is evident in Figure~\ref{fig:mono_illustration}, which corresponds to a deployment shown in Figure~\ref{fig:mono_deployment}.  Here, we run all components of PreFLMR in every server node: the logic is componentized but we collocate all the components even so.  As a result, nodes 0-3 have identical contents (as a reminder, in the case of PreFLMR A is the text encoder, B is the vision encoder, C performs cross-attention on matched pairs of (A,B) inputs, and D is the Colbert search).  The width of the boxes represents GPU memory in a active use, while the height of the boxes represents GPU computational activity.  We can see that while C, D and A are running the GPU is not fully used.  

In contrast, consider the multi-stage deployment of PreFLMR  illustrated in Figures~\ref{fig:micro_illustration} and \ref{fig:micro_deployment}. In this configuration, three servers  each run a single instance of step B, while server node 3 runs one instance each of steps C and A and three instances of step D, employing NVIDIA's multi-instance GPU (MIG) feature to split node 3's single GPU four ways.  The GPUs are much closer to fully loaded, and the aggregated capacity rises from 8 queries in the time period shown to 15.  Although today's {\sysname} requires some help from the developer to achieve this form of scaling, it should be possible to fully automate discovery of such opportunities \-- and even to dynamically vary the pattern over time.  Thinking ahead to enterprises with large numbers of ML pipelines and shared ML services, the technique will surely become important.

\subsubsection{Static Placement}
\label{sec:allocation}
To identify candidate deployment options, we ask which mappings accommodate the resource requirements as a function of the throughput levels achievable in each component.  This formulation yields a resource-constrained scheduling problem with replication and partitioning tradeoffs, solvable using Integer Linear Programming (ILP).

\noindent\textbf{Model and Parameterization.}

To model the system under resource constraints, we use available GPU resources and model performance profiles as inputs. The system comprises a fixed set of GPUs, denoted by $n \in \mathcal{N}$. Each GPU must be assigned one valid MIG layout from a set $\mathcal{L}$, where each layout is a multiset of MIG instance sizes that together sum to 24GB (e.g., $[6,6,6,6]$, $[12,6,6]$).

For each model $m \in \mathcal{M}$, we profile and store its latency $L_{m,c}$, throughput $T_{m,c}$, and memory usage $R_{m,c}$ when executed on a MIG instance of size $c \in \mathcal{C} = \{6, 12, 24\}$. These runtime profiles serve as input to the optimization problem.

The goal is to determine the number of replicas for each model and their assignment to MIG instances ($x_{m,n,c}$), as well as the selected MIG layout for each GPU ($y_{n,l}$).

\noindent\textbf{Constraints.}  
We enforce two main constraints: GPU layout validity and memory capacity.

\begin{itemize}[noitemsep, topsep=0pt]
    \item \textbf{MIG layout constraint:} Each GPU must be assigned exactly one MIG layout, expressed as $\sum_{l \in \mathcal{L}} y_{n,l} = 1 \ \forall n \in \mathcal{N}$.
    
    \item \textbf{Memory constraint:} A model replica can only be placed on a MIG instance of size $c$ if its memory requirement does not exceed $c$, i.e., $R_{m,c} \leq c$.
\end{itemize}

\noindent\textbf{Objective.}  
The objective is to maximize the minimum throughput across all workflow components. The throughput of a component $i$ is defined as:
% \begin{equation*}
$\text{Throughput}_i = \sum_{j \in \text{cluster}_i} T_{j}$,
% \end{equation*}
where $T_j$ is the throughput of model replica $j$ assigned to that component. The overall objective is:
% \begin{equation*}
$\max \left( \min_{i} \text{Throughput}_i \right)$.
% \end{equation*}
When additional resources are available, the optimization proceeds to maximize the second-lowest component throughput, and so on in a lexicographic manner.

\noindent\textbf{Implementation.}  
We solved this ILP problem using the Gurobi optimizer, yielding the configurations shown in our multi-stage figures and experiments.

\section{Experiments}
\label{sec:experiments}
In this section we experimentally assess the methodologies described in Section ~\ref{sec:challenges}: microservice deployment, data caching, compute collocation, opportunistic batching, and efficient stage-to-stage handoffs.

\subsection{Experiment Environment}
We conduct our experiments on a cluster of 12 D7525 servers provisioned through CloudLab. Each node is equipped with two 16-core AMD EPYC 7302 processors, 128 GB of ECC memory, and a single NVIDIA A30 GPU (24 GB, Ampere architecture). The nodes are interconnected via dual-port Mellanox ConnectX-6 DX 100 Gb NICs, with one port configured at 200 Gb/s. We configure RoCE (RDMA over Converged Ethernet) over this link to enable RDMA-based communication between nodes during the experiments.

\subsection{Datasets}
Our experiments on PreFLMR (pipeline 1) run on the EVQA~\cite{lyu2023semantic} portion of the Multi-Task Multi-Modal Knowledge Retrieval Benchmark dataset.  Each query consists of one question, one instruction, one image and a collection of contextual passages.  Images come from two sources, the google-landmarks dataset~\cite{weyand2020GLDv2} and iNaturalist~\cite{van2018inaturalist}. The model generates one query embedding for each data entry using the question and instruction as the text input, and the image as a vision input, and the pipeline completes when the ColBERT RAG search returns the most relevant context passages, using a prebuilt inverted flat product quantization (IVFPQ) index.  Experiments on AudioQuery (pipeline2) draw data from MS MARCO~\cite{nguyen2016ms} version 2.1.  We generated audio renderings of each question using the Tacotron2 TTS model~\cite{ttsTacotron2}, and pretrained an IVFPQ index on the underlying document corpus for the RAG lookup, which is performed using FAISS ~\cite{faiss}. 

\subsection{Metrics}

Our experiments employ the following metrics:

\noindent\textbf{Latency}: End-to-end, from when a query enters the serving system until the response is complete.

\noindent\textbf{SLO miss rate}: The workflows we evaluate are interactive and latency-sensitive, each with defined latency service-level objectives (SLOs). The SLO miss rate measures the fraction of queries whose end-to-end latency exceeds a given SLO target.
% Same metrics are used by ML serving sntystems such as ~\cite{inferline, INFaaS}.

\noindent\textbf{Throughput}: Queries per second (QPS) that can be sustained without backlog.

% \noindent\textbf{GRACT} (\textit{graphics engine activity}).  The fraction of time GPU compute was active.  GRACT normally also includes graphics compute, but our ML tasks don't use those capabilities.  The GRACT metric is a common proxy for more detailed GPU utilization measurements, given the lack of options for finer-grain instrumentation ~\cite{MIGPerf,profile_training_euromlsys}.

Our goal, then, is to ask whether {\sysname} is an effective platform for maximizing throughput while minimizing latencies, meeting the SLO targets and whether it leverages the hardware effectively.  Implicit is the assumption that this mix optimizes cost of owning and operation the ML service while respecting application-specific SLOs.

\subsection{Serving Paradigms}

Our coarse-grained experiments look at TorchServe~\cite{torchserve}, Ray Serve~\cite{ray,ray_serve_docs} and {\sysname}.  TorchServe was a majority solution for many years and remains popular, but has poor throughput in many configurations. Ray Serve is a far more performant and widely used option at the time of this writing.  We do include TorchServe in one experiment, but our detailed microbenchmarks focus purely on Ray Serve and {\sysname}.  Ray Serve includes an autoscaler, but we found it to be considerably less effective than manual configuration, so both {\sysname} and Ray Serve were identically hand-configured, ensuring a fair comparison.  

\begin{figure}[ht]
 \centering
 \includegraphics[width=0.41\textwidth]{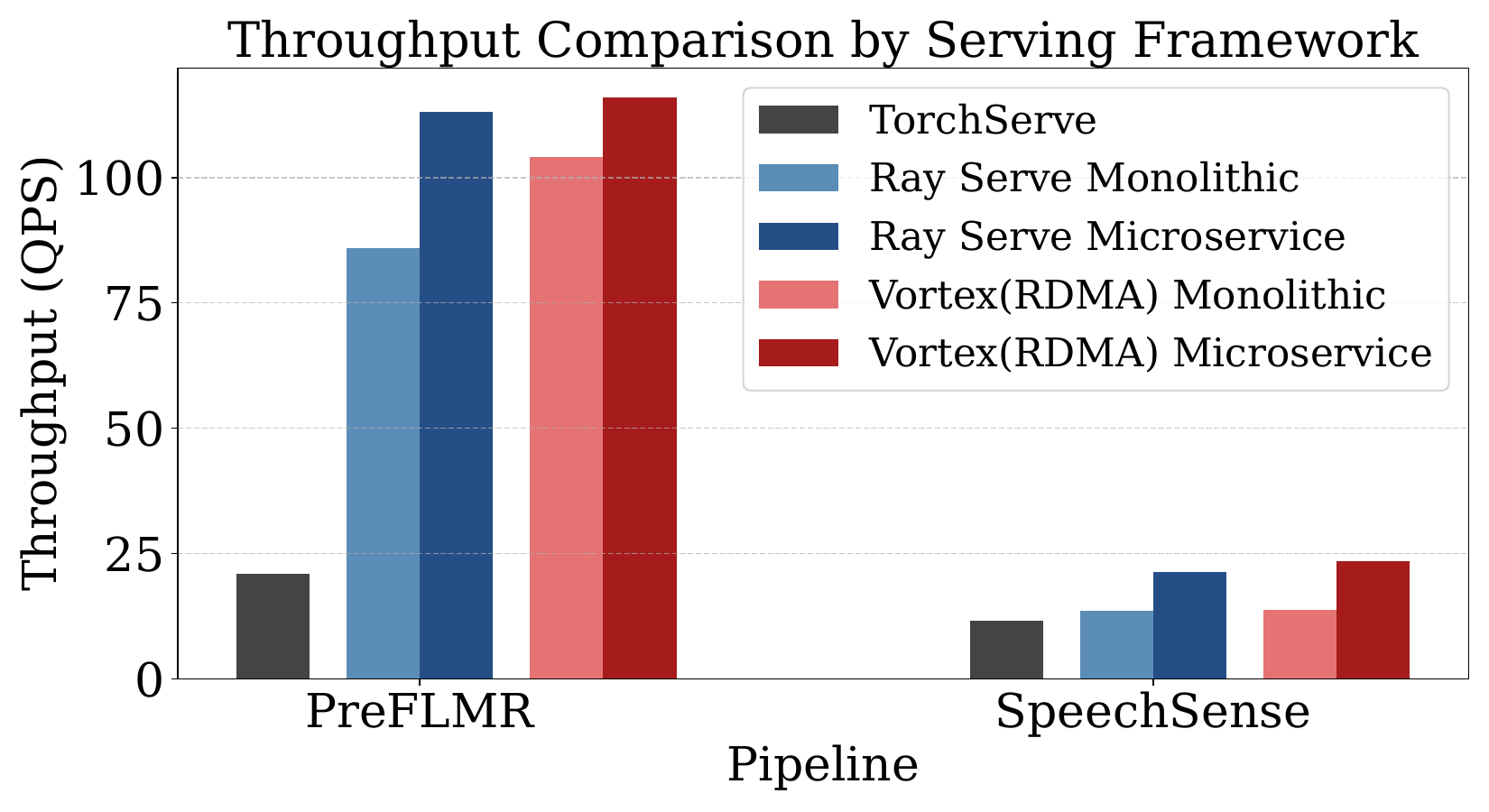}
 \vspace{-1.0em} 
 \caption{TorchServe, Ray Serve and {\sysname} on a 4-node cluster.}
 \label{fig:achievable_throughputs_comparison}
\end{figure}

% Torch Serve supports building the pipelines in DAG structure, but it is limited to single host, not across multiple hosts or processes. Therefore, we deploy Torch Serve in its native framework in monolithic paradigm, and scale it by replicating the whole pipeline to multiple servers. Ray Serve supports configurable distributed ML serving pattern where different stage could be allocated to the same or different processes.

\begin{figure}[H]
 \centering
 \includegraphics[width=0.45\textwidth]{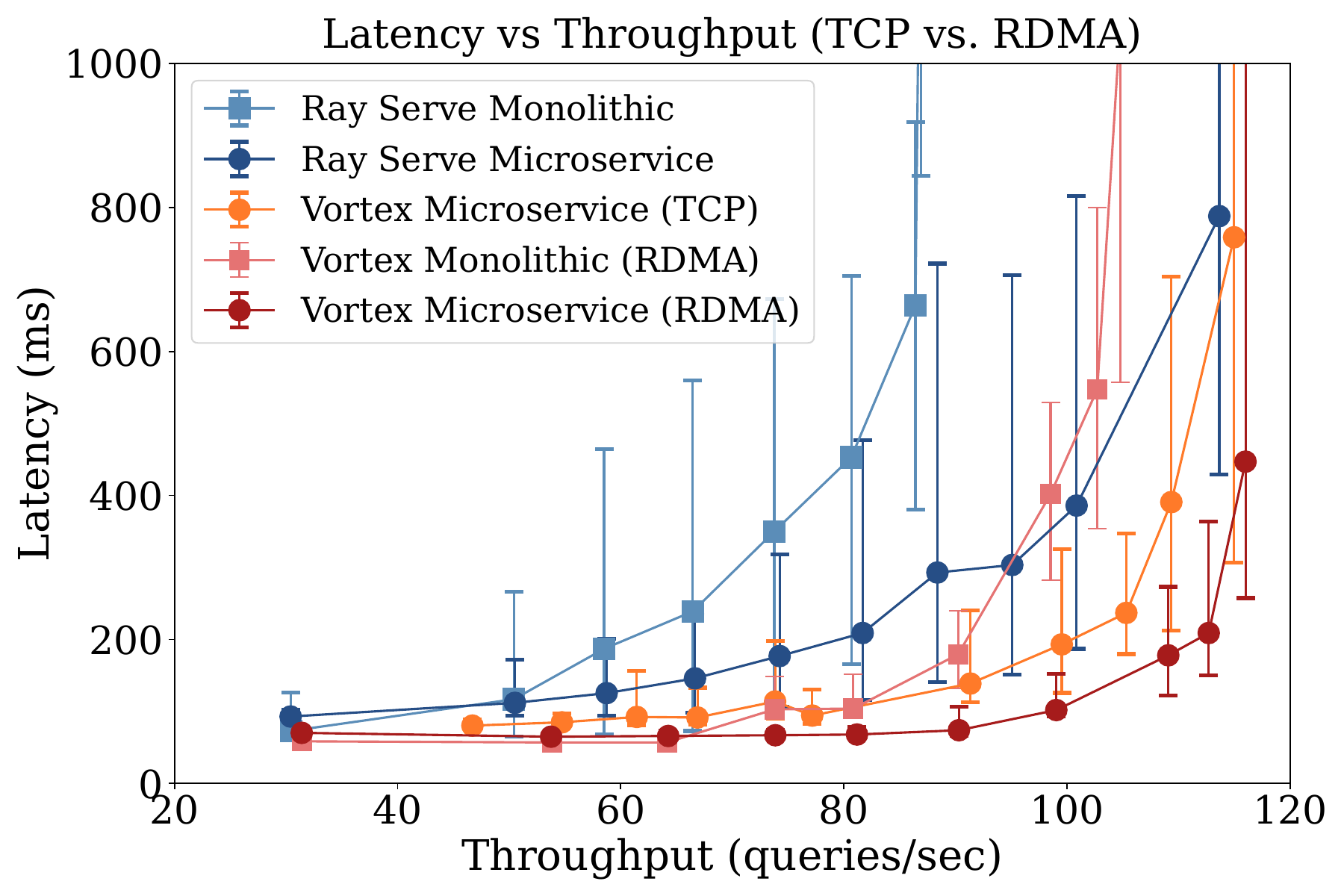}
 \vspace{-1.0em} 
 \caption{ 
    4-node PreFLMR performance comparison across serving paradigms and network protocols.  The x-axis gives the presented load, and the y-axis median latencies with 95\% and 5\% bars.
    We consider Ray Serve (monolithic and microservice) with {\sysname} (monolithic and microservice), with and without RDMA. Note that Ray Serve does not currently leverage RDMA.  }
 \label{fig:latency_vs_throughput_all_configurations}
\end{figure}

\begin{figure*}[ht]
    \centering
    \begin{subfigure}[b]{0.48\textwidth}
        \centering
        \includegraphics[width=0.85\textwidth]{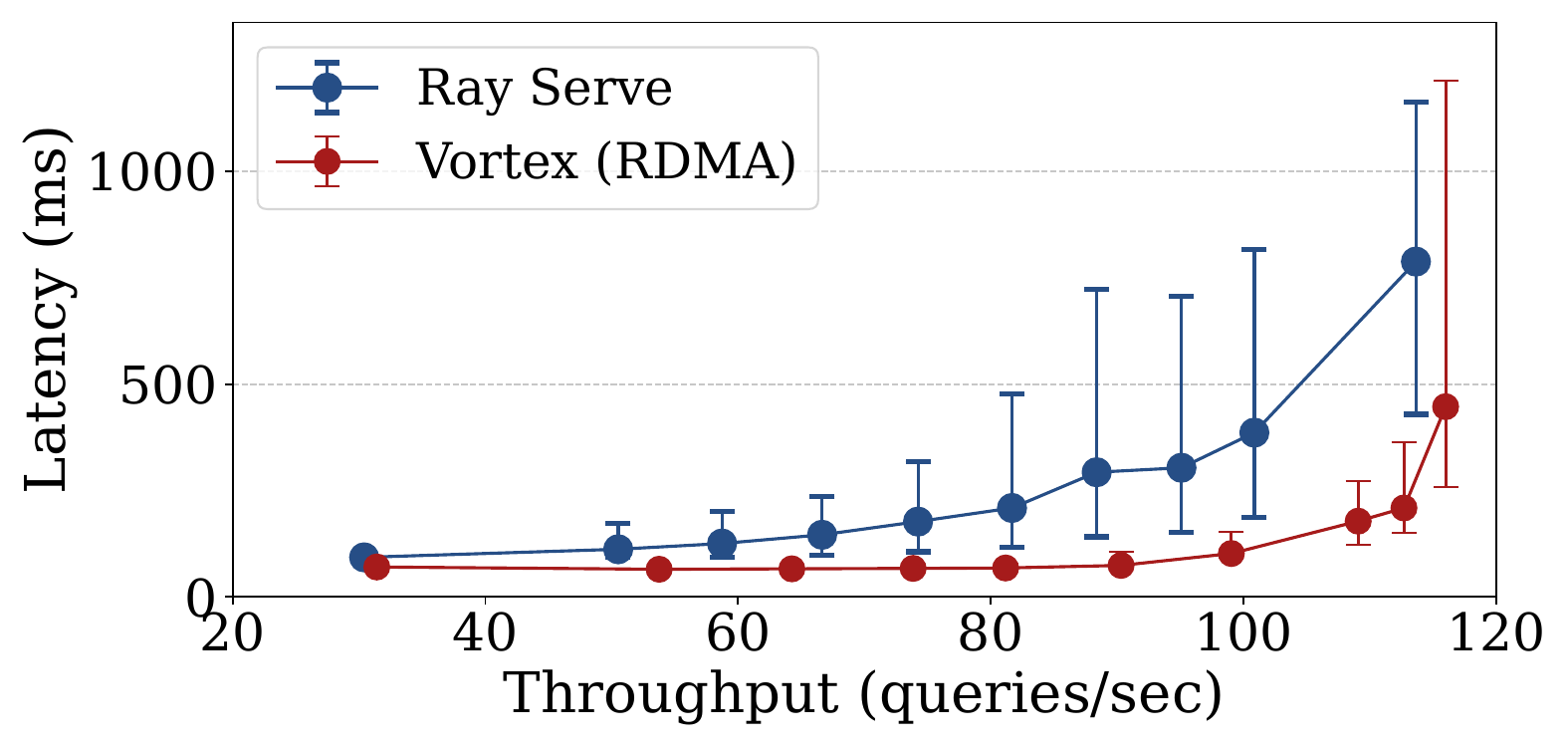}
        \caption{\centering End-to-end latency vs. throughput for \textbf{PreFLMR} pipeline}
        \label{fig:preflmr_lat_throughput}
    \end{subfigure}
    \begin{subfigure}[b]{0.48\textwidth}
        \centering
        \includegraphics[width=0.85\textwidth]{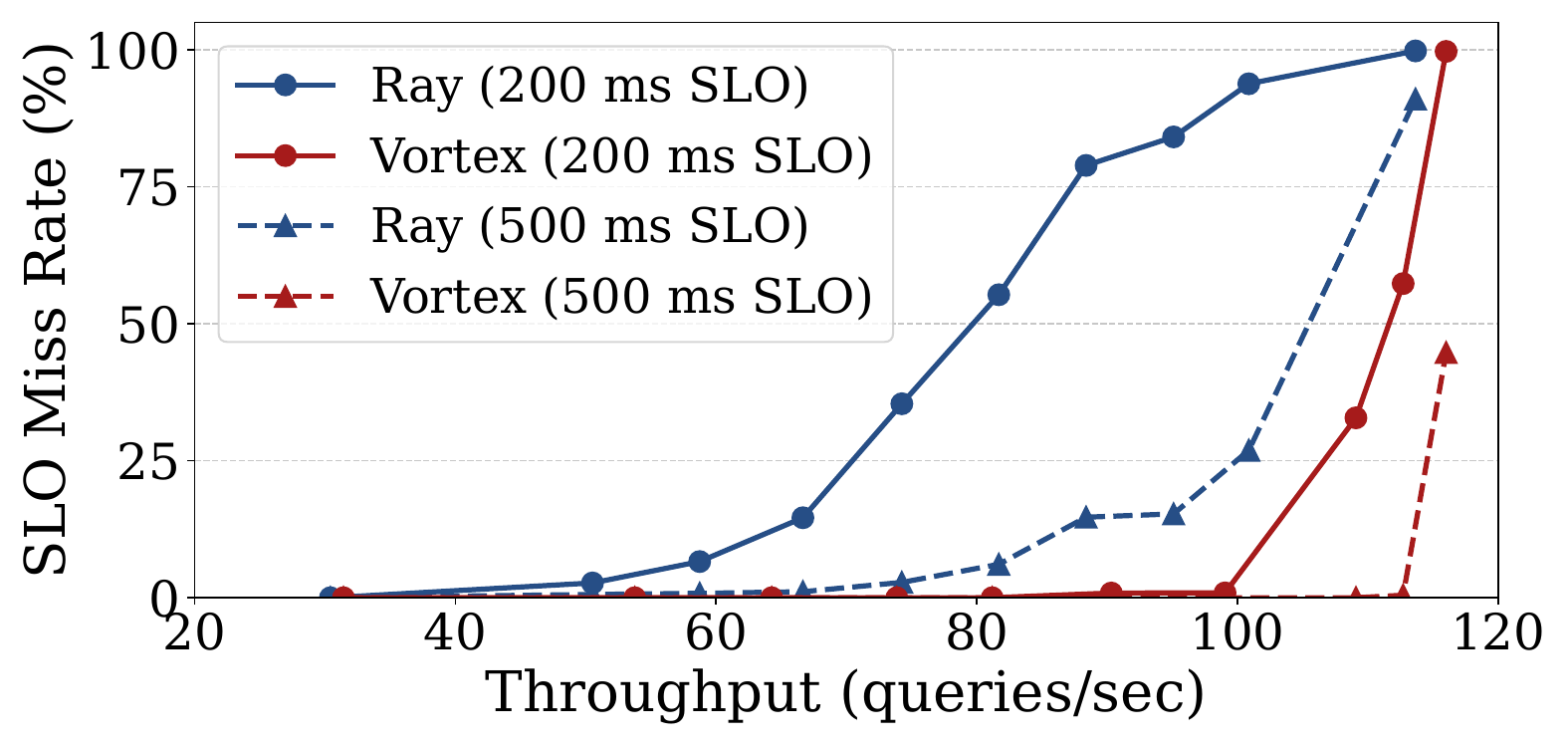}
        \caption{\centering SLO miss rate vs. throughput for \textbf{PreFLMR} pipeline}
        \label{fig:preflmr_slo_throughput}
    \end{subfigure}

    % \vspace{0.5em}
    \centering
    \begin{subfigure}[b]{0.48\textwidth}
        \centering
        \includegraphics[width=0.85\textwidth]{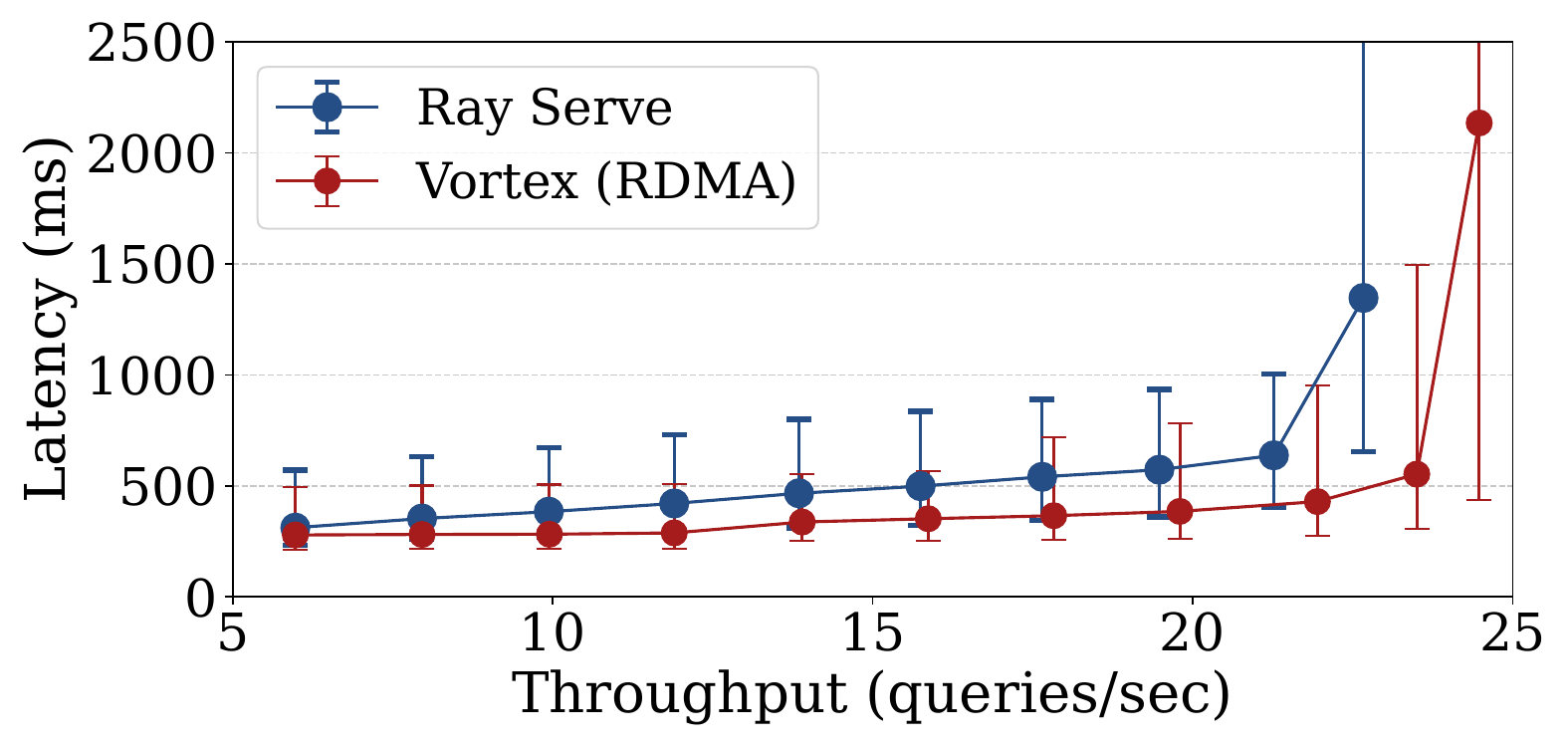}
        \caption{\centering End-to-end latency vs. throughput for \textbf{AudioQuery} pipeline}
        \label{fig:AudioQuery_lat_throughput}
    \end{subfigure}
    \begin{subfigure}[b]{0.48\textwidth}
        \centering
        \includegraphics[width=0.85\textwidth]{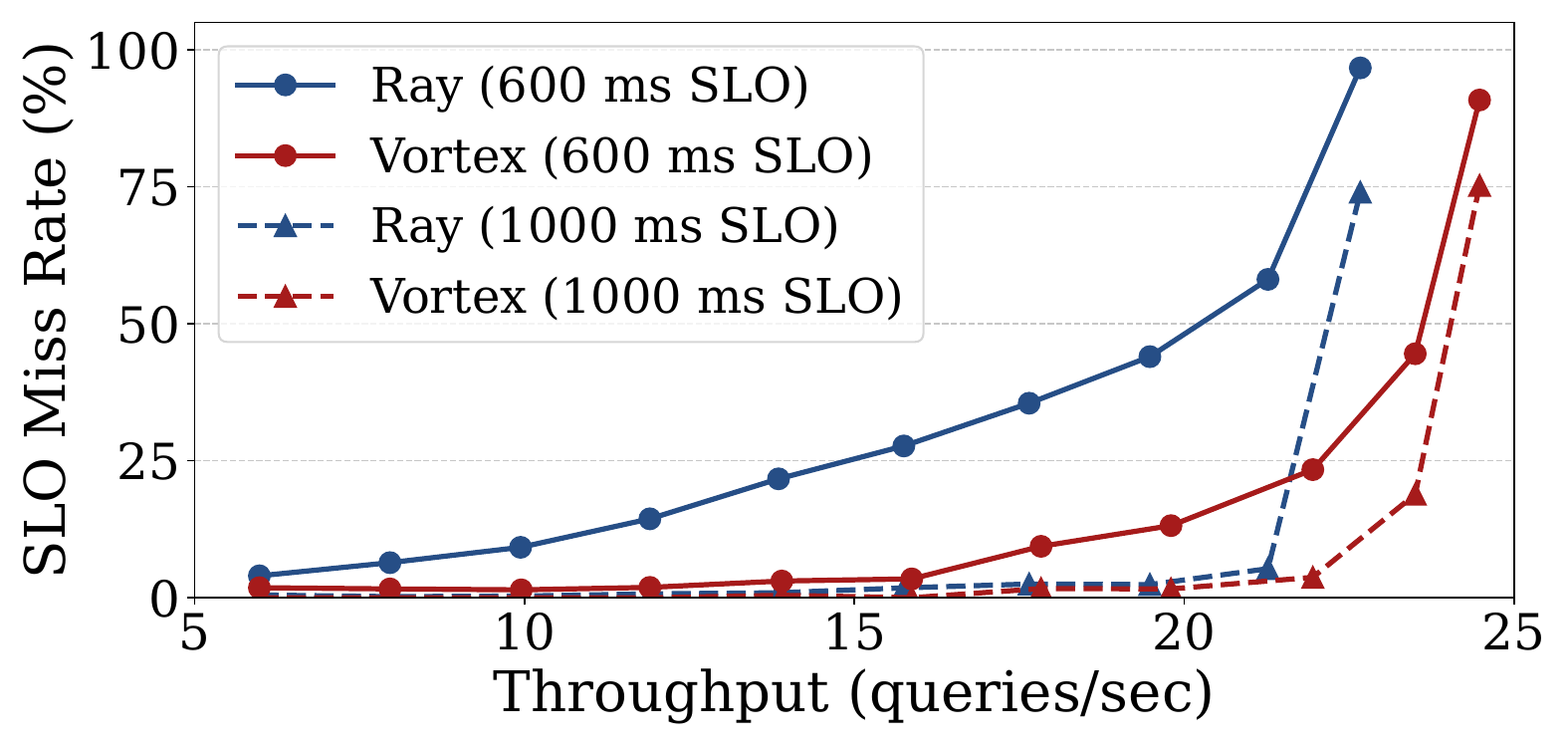}
        \caption{\centering SLO miss rate vs. throughput for \textbf{AudioQuery} pipeline}
        \label{fig:AudioQuery_SLO_throughput}
    \end{subfigure}

    \vspace{-1.0em} 
    \caption{End-to-end latency of \textbf{PreFLMR} and \textbf{AudioQuery} on 4 nodes. 
    The plots on the left (a,c) report latency values at each throughput level, while the plots on the right (b,d) show the corresponding SLO miss rates. Note that Ray Serve does not leverage RDMA.}
    \label{fig:ray_vs_typhoon_deployment}
\end{figure*}

\subsubsection{Monolithic vs. Microservice}
\label{sec:exp_mono_vs_micro}

Many production inference and knowledge retrieval servers run as monolithic deployments. Our pipelines leverage GPU acceleration but inference servers are generally less heavily configured than model training systems.   Accordingly, our monolithic setup scales by replicating the entire deployment to however many machines are needed, as demonstrated in Figure~\ref{fig:mono_deployment} for the PreFLMR workflow. Our microservice configurations are determined using the methodologies outlined in Section~\ref{sec:challenges}, including optimized resource partitioning, model batching and placement. 
Recall that Figure~\ref{fig:micro_deployment} illustrates a possible microservices deployment of this same pipeline. This was the basis of both the Ray Serve and {\sysname} deployment and exactly matches what both systems are optimized for. TorchServe only permits monolithic deployments, but we did everything we could to optimize within the limits of that platform. 

Figure~\ref{fig:achievable_throughputs_comparison} compares the best achievable throughput across all three serving frameworks on a 4-node cluster running both the PreFLMR and AudioQuery pipelines.
Ray Serve and {\sysname} outperform TorchServe, achieving \textbf{1.8$\times$ to 5.5$\times$} higher throughput, primarily because of TorchServe's data transfer/deserialization overheads. 
For both Ray Serve and {\sysname}, the microservice deployment achieves higher throughput than the monolithic setup, with improvements ranging from \textbf{10\% to 56\%} across the PreFLMR and AudioQuery pipelines.

Recall from Figure~\ref{fig:micro_deployment} that our favored microservices deployments for PreFLMR and AudioQuery involve  data transfers between stages: an overhead not seen in a monolithic deployment, and one capable of bringing significant overheads, especially given that the output size of the vision encoder is relatively large (10-20MB) in the PreFLMR pipeline.  This effect is evident in a subtle way: the very best achievable latencies for the microservice deployments on Ray Serve and {\sysname} are slightly inferior to best monolithic latencies on the same machines.  However, a user would pay a steep price to gain a very small improvement in latency: to achieve those best-possible numbers, they would be limited to very low throughput (hence, a high per-query cost). 

As seen in Figure~\ref{fig:latency_vs_throughput_all_configurations}  the latency for monolithic deployment degrades more sharply than that for the microservice deployments as system load increases for both Ray Serve and {\sysname}. Ray Serve’s microservice deployment achieves a median latency approximately \textbf{1/3} that of its monolithic counterpart, while {\sysname} achieves a latency reduction of \textbf{1/2} to \textbf{1/4} at the same system throughput.

The bottom line for our effort centers on the SLOs that services can offer, and the tradeoffs between SLO and throughput.  This question is investigated in 
Figure \ref{fig:ray_vs_typhoon_deployment}, which compares the latency and SLO performance of Ray Serve and Typhoon, illustrating how latency distributions impact SLO attainment.  Notice that here we are comparing a Ray Serve configuration that runs on TCP (because Ray Serve does not currently support RDMA) with a {\sysname} configuration on RDMA.  For the PreFLMR pipeline in the configurations we explored the SLO miss rate of Ray Serve is higher than that of Typhoon (Figure \ref{fig:preflmr_slo_throughput}). Both systems achieve a 0\% miss rate at low request rates, but the gap widens as load increases. At a throughput of 100 QPS, Ray Serve’s SLO miss rate reaches \textbf{93.8\%} for a 200 ms SLO target and \textbf{26.9\%} for a 500 ms target, while Typhoon maintains just \textbf{0.89\%} and \textbf{0\%}, respectively. Figure~\ref{fig:AudioQuery_SLO_throughput} shows a similar trend in the AudioQuery pipeline.  Had we included {\sysname} on TCP, the results would have be similar but the gap between Ray Serve and {\sysname} would be roughly halved, representing the benefit {\sysname} gains by leveraging RDMA. 
% for AudioQuery, at 19.48 QPS, Ray Serve’s miss rate is 44 \% for the 600 ms SLO and 2.4 \% for the 1000 ms SLO, whereas Typhoon at a comparable throughput of 19.8 QPS records only 13.16 \% and 1.61 \%, respectively.
\captionsetup[subfigure]{skip=0pt, aboveskip=-1pt, belowskip=-1pt}
\begin{figure}[ht]
 \centering
 \begin{subfigure}[b]{0.44\textwidth}
     \centering
     \includegraphics[width=\textwidth]{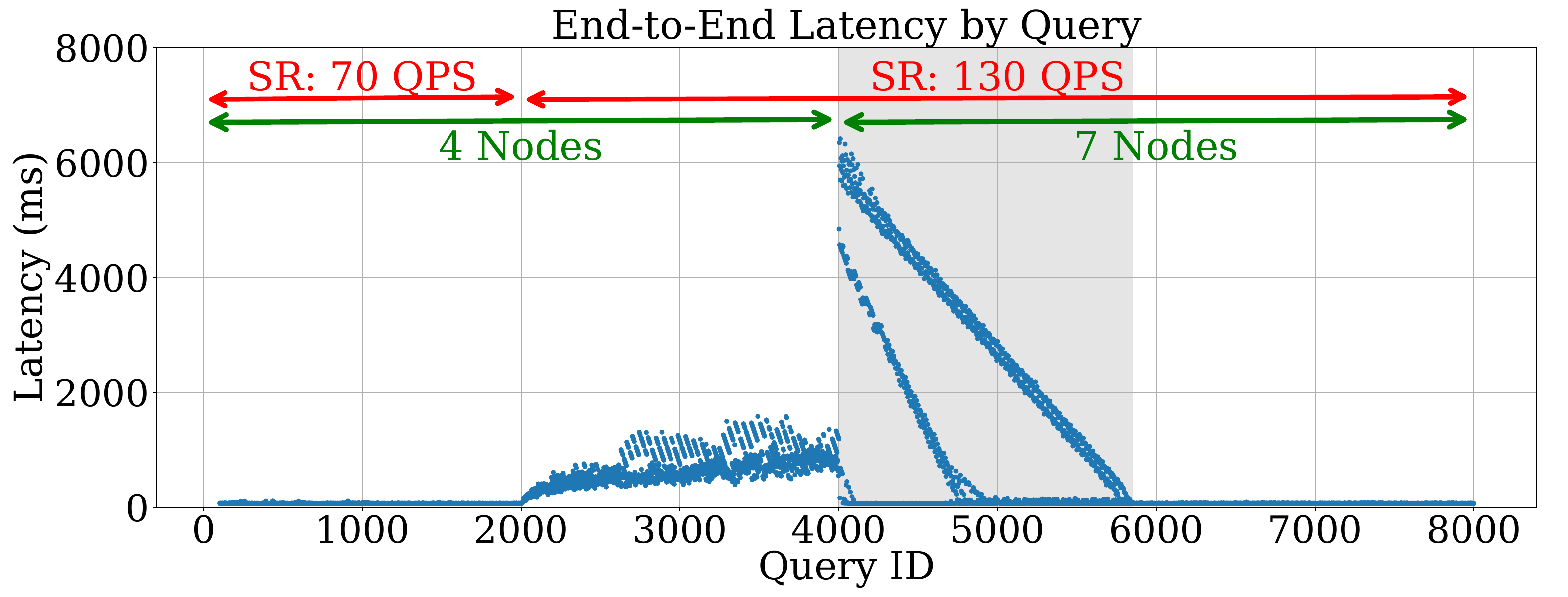}
     \caption{Elastic resizing without preloading models.}
     \label{fig:cold_start}
 \end{subfigure}%
 \hfill%
 \begin{subfigure}[b]{0.44\textwidth}
 \centering
 \includegraphics[width=\textwidth]{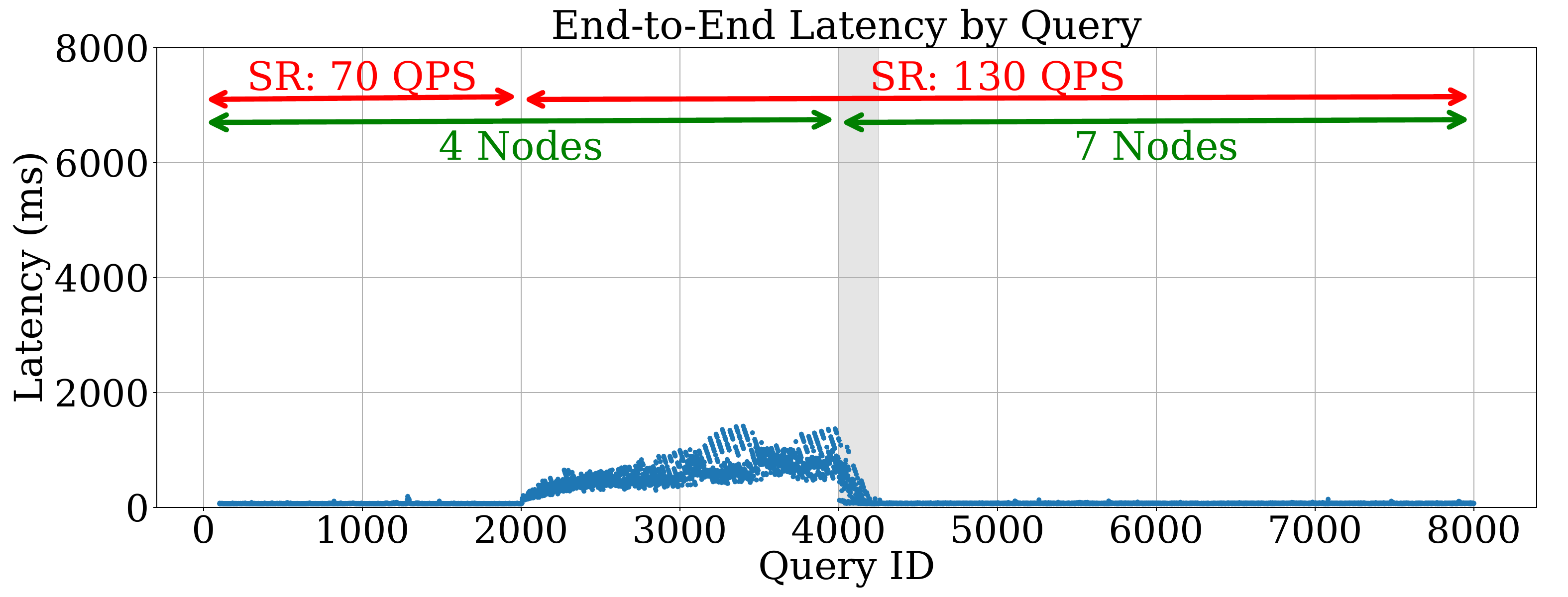}
 \caption{Benefit of preloading models in GPU memory.}
 \label{fig:with_warmup}
 \end{subfigure}
 \vspace{-1.0em} 
 \caption{After 2000 queries on a \textbf{{\sysname}} 4-node deployment at  70 QPS the rate rises to 130 QPS.  At query 4000, 3 instances are added.  Preloading the ML model and other dependent objects avoids a latency spike.}
 \label{fig:dynamic_send_rate}
\end{figure} 

\subsubsection{Sustaining SLOs across resizing events}
\label{sec:cold_start}
Recall the concerns about resizing expressed in Section~\ref{sec:elasticity}: if we suddenly increase the pool of workers to handle a load surge, the ML service might stall while loading models, causing a burst of SLO misses.   This effect is evident in Figure~\ref{fig:cold_start}, where a load surge from 70 to 130 QPS forces the scheduler to increase the number of servers from 4 to 7 starting at query id 4000; the stall that occurs results in inflated latencies for tasks sent to the new server and also creates a cascade of delays.  High and variable latencies are evident until query 6000.   Far better is the anticipatory approach, as shown in Figure~\ref{fig:with_warmup}.  Here, the load surge is detected early and {\sysname} initiates model preloading in anticipation of possible resizing.  Preloading does bring costs, but the overhead is insignificant.  Both the worker stall and the SLO misses are avoided.  

% When the send rate is high, like in the example plot, scaling without warmup leads to 1115 queries end-to-end latency exceeding 1000ms in this , comparing to with warmup 173 points exceeding 1000ms.

\subsection{Efficient stage-to-stage handoff effects}
\label{sec:exp_batching}

\begin{figure}[htp]
\begin{subfigure}[b]{0.9\columnwidth}
    % \centering
 \includegraphics[width=\columnwidth]{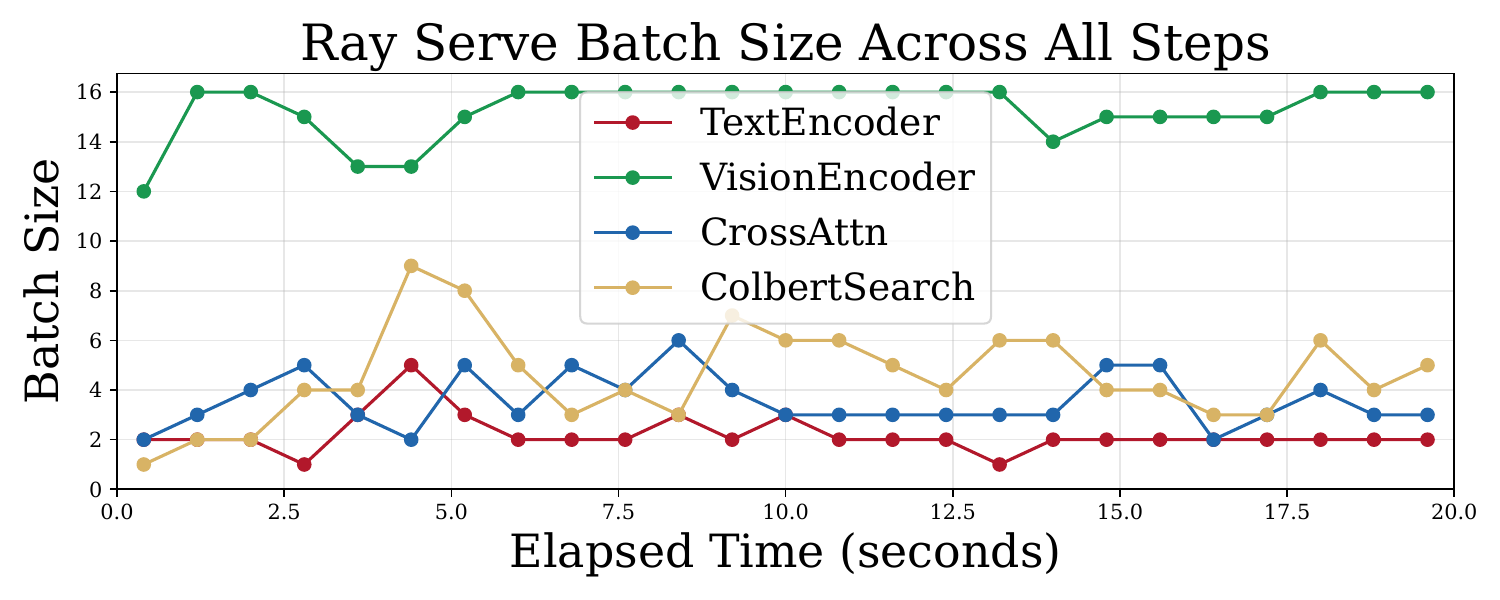}
 \label{fig:ray_batch_size}
\end{subfigure}
  \vspace{-0.7em}  % tighten vertical gap between the two plots
 \begin{subfigure}[b]{0.9\columnwidth}
     % \centering
 \includegraphics[width=\columnwidth]{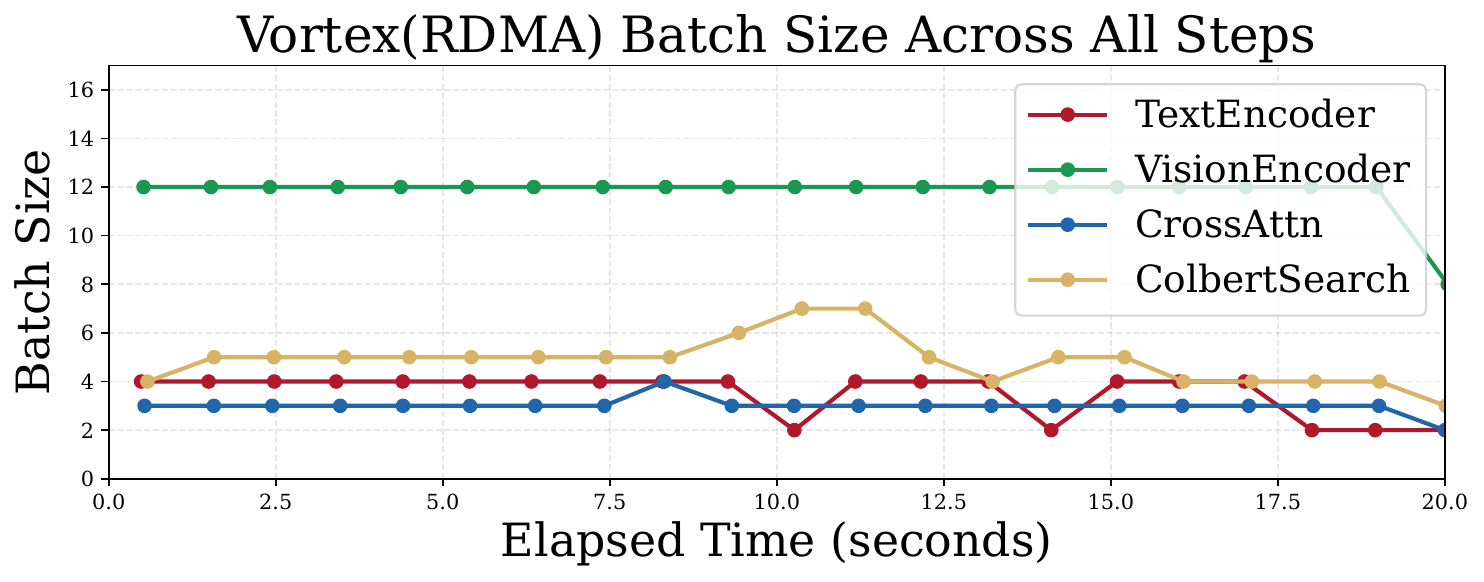}
 \label{fig:vortex_batch_size}
 \end{subfigure}
 \vspace{-1.5em} 
 \caption{Median batch sizes for the components of PreFLMR at ~214qps (high load).}
 \label{fig:batch_size_across_all_steps}
   % \vspace{-0.3em}  
\end{figure}

% Rewrote with numbers and rephrased some points
{\sysname}’s architecture, combined with its use of RDMA for inter-node communication when that option is available leads to improved end-to-end latency.  In any ML pipeline, some degree of end-to-end latency variability is inevitable because queries stream in at unpredictable rates, and different stages settle into different individualized batch sizes (with upper bounds configured to minimize SLO misses).  The opportunistic batching used by both systems allows them to catch up when backlogs arise, improving throughput.  Yet Figure~\ref{fig:batch_size_across_all_steps} shows a substantially reduced degree of variability for {\sysname}, reflecting the heavy optimization of the {\sysname} stage-to-stage handoff and its leverage of RDMA communication paths when available: even though both graphs reflect the same rate of incoming queries, {\sysname} is less prone to internal backlogs.  

Ray Serve is limited in part by the lower bandwidth of TCP, but also by some instances in which Ray's server selection seems to have used stale load information (dispatching a request to a busy server rather than a lightly-loaded one).  

%\jamal{The stage-to-stage handoff optimizations explain some of the performance increase, but not all of it. The rest comes from Ray not routing requests very well. E.g. Even if there's a free colbertsearch node available, sometimes Ray sends a request to a busy one anyways. Should we mention this?}

\begin{figure}[t]
    % \captionsetup[subfigure]{skip=0pt, aboveskip=-2pt, belowskip=-3pt}
     \begin{minipage}[b]{0.42\textwidth}
         \centering
         \includegraphics[width=\textwidth]{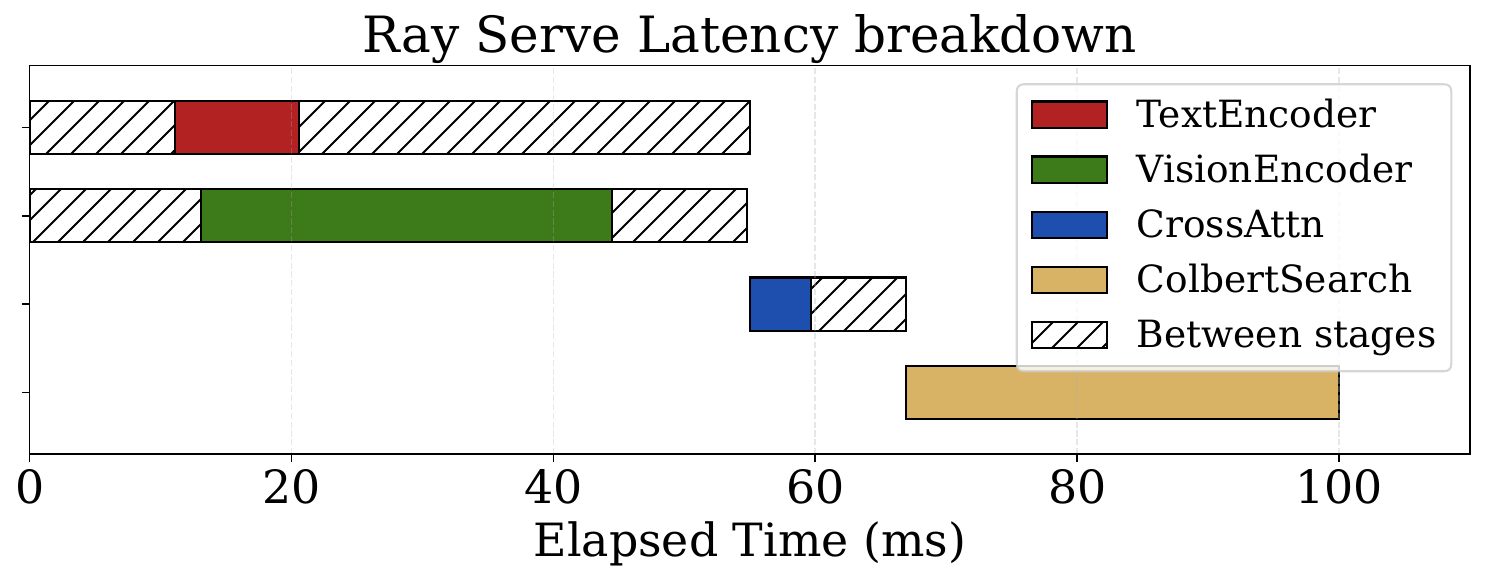}
         % \caption{PreFLMR pipeline latency breakdown for Ray.}
         \label{fig:latency_breakdown_ppl2}
     \end{minipage}
     % \hfill
     \hspace{-20pt}
     
     \begin{minipage}[b]{0.42\textwidth}
         \centering
         \includegraphics[width=\textwidth]{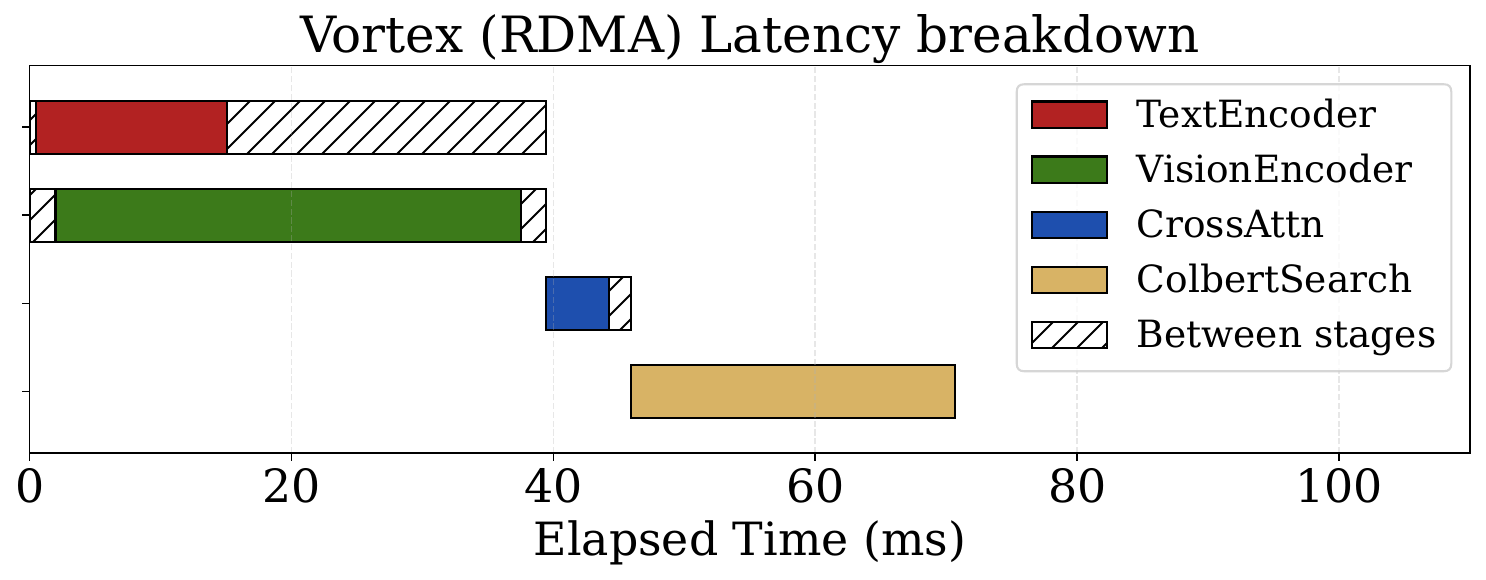}
         % \includegraphics[width=\textwidth]{plots/sys_perf/typhoon/latency_breakdown.pdf}
         % \caption{PreFLMR pipeline latency breakdown for {\sysname}.}
         \label{fig:latency_breakdown_ppl1}
     \end{minipage}
     \vspace{-2.0em} 
\caption{Latency Breakdown by the average latencies of all components in PreFLMR, at 32qps (low load) on a 4-node cluster.}
\label{fig:latency_breakdown}
\end{figure}

Figure~\ref{fig:latency_breakdown} breaks end-to-end latency down by execution component and stage-to-stage handoff delays. Under low system load queuing-delays do not arise within the pipeline and {\sysname} achieves an average end-to-end latency of 70ms.  Ray Serve achieves 93ms: \textbf{33\% } higher.  Notice that Colbert Search is slightly slower on Ray Serve: possibly a NUMA memory access or locking effect. {\sysname}’s stage-to-stage data transfers complete in under 2ms, whereas Ray Serve, using TCP rather than RDMA, requires 5–13ms. {\sysname} maintains that speed even for large intermediary results such as the outputs of the Vision Encoder and Cross-Attention components. The key takeaway is that for steady high throughput with low latency, it is important to minimize copying on the critical path and to leverage RDMA.
%With all of our optimizations enabled, {\sysname} was able to achieve higher throughput while maintaining low latency at high loads. This suggests that {\sysname} would be well-suited to environments where low latency is critical. Ray Serve implements many of the optimizations we detailed, including opportunistic batching, and compute collocation, but does not implement some of the stage-to-stage handoff optimizations, and does not support RDMA and data dependencies as {\sysname} does. 

When we discussed Figure~\ref{fig:latency_vs_throughput_all_configurations} we noted that  Ray Serve's latency is \textbf{1.3x to 3.9x} higher than {\sysname}-RDMA at the same system load. 
When we configure {\sysname} to use TCP rather than RDMA, throughput drops by \textbf{1.7\%} and latency increases by \textbf{1.23x to 1.9x}.
We further assess the impact of fast handoffs by comparing {\sysname}’s performance over TCP with that of Ray Serve, both under microservice deployment. Thus even without RDMA, {\sysname}-TCP achieves higher throughput and lower latency than Ray Serve, highlighting that the optimizations we employed to leverage RDMA (such as zero-copying end-to-end data paths and avoidance of locks on critical paths) are beneficial even when RDMA is not available. It will be fascinating to see how leveraging RDMA in Ray Serve impacts (reported by AnyScale as a planned feature) impacts that system's TCP performance.

\captionsetup[subfigure]{skip=0pt, aboveskip=-2pt, belowskip=-3pt}

\section{Related Work}

\noindent\textbf{SLO-Focused ML Services}
The Tail at Scale~\cite{dean2013tail} launched the modern focus on SLOs.  Examples of other systems that explore SLO-oriented ML serving include include Autothrottle~\cite{wang2024autothrottle}, INFaaS~\cite{romero2021infaas}, Clipper~\cite{crankshaw2017clipper}, Inferline~\cite{inferline}, and Sinan~\cite{sinan-asplos2021}. Our work is unusual in focusing on componentized ML pipelines within which RDMA-friendly critical paths, efficient stage-to-stage handoffs, opportunistic batching and anticipatory provisioning for autoscaling enable tighter SLOs with higher throughput. 
Inferline~\cite{inferline} is most relevant to the scheduling component of this work. It addresses autoscaling by predicting changes in request arrival rates to adjust the number of provisioned machines and the batch size. Typhoon provides a more general execution framework that allows such schedulers to operate across multi-stage, distributed ML pipelines. % this part has some overlaps with ML serving platform section, should we combine these two?

% We look closer into the scaling transition effects on the requests' latencies and SLO misses, as detailed in Section~\ref{sec:exp_batching}.

\noindent\textbf{LLM Inference.}
% \emph{ML Inference.} 
LLM serving is receiving growing attention and extensive research has been conducted about optimizing individual model inference from both execution~\cite{orca, apparate,distServe,sarathi-serve} and memory management~\cite{pagedAttention,flexgen,jenga,distServe, aqua} perspectives.  
% Orca ~\cite{orca} schedules and manages the execution at iteration-level granularity for Transformer models. DistServe~\cite{distServe} and SarathiServe~\cite{sarathi-serve} optimize the LLM decoding and prefill phases. Apparate~\cite{apparate} explores the early-exit strategies in ML serving. 
% In terms of memory management, PagedAttention~\cite{pagedAttention} uses virtual memory and paging to enable efficient KV cache management; FlexGen~\cite{flexgen} offloads tensors (such as weights, activations, and key-value (KV) cache to CPU and disk), and compresses the model weights and attention cache, enabling LLM inference under limited GPU memory; and Jenga~\cite{jenga} addresses memory fragmentation in heterogeneous embedding workloads. Aqua~\cite{aqua} reduces paging overheads via  preemptive scheduling with cross-GPU memory management.  Notice that all of these works focus on optimizing individual LLM inference.  
In contrast, {\sysname} targets the deployment of end-to-end pipelines that integrate one or more such models. It addresses the challenges of handling upstream multimodal inputs and downstream tasks, and focuses specifically on optimized deployment for microservice layouts that span multiple hosts.

\noindent\textbf{LLM Pipelines.}
Prior efforts have also considered componentized pipelines in the context of LLM serving and agent workflows~\cite{ayo,autellix}.
Ayo~\cite{ayo} introduces a static, fine-grained dataflow graph to represent end-to-end LLM applications, optimizing execution through existing distributed engines such as Ray and vLLM. {\sysname} adopts a similar componentization approach but with more emphasis on leveraging RDMA networking and GPU sharing, both of which were shown to enhance performance.
Autellix ~\cite{autellix} formalizes agentic programs as dynamic DAGs, then schedules by preempting and prioritizing LLM calls, whereas {\sysname}'s system-level optimizations benefit runtime behavior of such dynamic DAG structures.  Combining the two approaches could be a worthwhile topic for future exploration. 

% Several commercial frameworks, including Haystack~\cite{haystack}, AutoGen~\cite{autogen,autogen_git}, LlamaIndex~\cite{llamaindex}, LangChain~\cite{langchain}, and LangGraph~\cite{langgraph}, facilitate the development of LLM-based applications and agentic workflows. However, these frameworks focus more on the LLM-based application development experience and addressing rhe challenge of interconnecting complex components to support a wide range of possible workflow patterns.  In contrast, our effort is more focused on runtime performance optimizations arising in distributed deployments.

\noindent\textbf{ML Serving platforms.}
There are a number of existing systems and products for general-purpose distributed ML serving~\cite{dynamo,nvidia_triton,ray_serve_docs,deepspeed, torchserve}.  
NVIDIA distributed serving framework Dynamo~\cite{dynamo} and Triton~\cite{nvidia_triton} are competitive commercial products that supports distributed ML inference. DeepSpeed~\cite{deepspeed} also supports distributed execution but is primarily designed for large-scale model training rather than inference.  Among systems that have received substantial research attention (some are also products), many target back-end training settings, such as Ray  ~\cite{ray,rayData}. Clipper~\cite{ clipper17}, AlpaServe~\cite{li2023alpaserve} and Nexus~\cite{shen2019nexus} place primary emphasis on scheduling and optimizing for traditional Deep Learning inference. Ray Serve~\cite{ray,ray_serve_docs} is more directly comparable to {\sysname}, and is notable for high performance and the ease with which it can be used for deploying general-purpose complex ML pipelines across distributed machines. Other related research includes Ray Data~\cite{rayData}, which focuses on data processing in ML inference and training for fault-tolerant heterogeneous execution. In our experimental section, we showed that Ray Serve is highly effective and can achieve the same throughput and hardware utilization as {\sysname} with manual configuration, but that {\sysname} has lower and steadier latencies.

%{\sysname} reference from the microservice techniques introduced in these papers, but focuses on the special characteristics brought by multimodality knowledge retrieval pipelines, with recent techniques such as GPU partitioning and network optimization to demonstrate how this new type of workflows if decomposed into DAG structure could benefit from these microservice techniques.

\noindent\textbf{Vector Search.}
Vector search is one of the core components in knowledge retrieval pipelines, including the two we used as running examples. There are numerous commercial or open-sourced projects that are built for vector search, including Pinecone~\cite{pinecone}, TimescaleDB~\cite{timescaledb}, Weaviate~\cite{weaviate}, pgvector~\cite{pgvector}.
Research papers on optimizing query vector search, RUMMY~\cite{fastVectorQueryPrococessing}, FAISS~\cite{faiss}, Colbert~\cite{colbert_serve}, DiskANN~\cite{diskAnn}, Parlay ANN~\cite{parlayAnn} explore algorithm and systems optimization to run nearest neighbor search efficiently.
{\sysname} considers the vector search as one of the stages in the pipelines in its end-to-end pipelines, and optimizes the whole pipeline in the context of distributed and multimodal execution. The pipelines we considered did include vector search stages (Colbert Search in the case of PreFMLR and FAISS in the case of AudioQuery).  However, we believe more could be done: the kinds of optimization and techniques introduced in these prior works could be integrated into the vector search stage in {\sysname}, and have the potential to further improve its performance.

% For the downstream tasks that we didn't focus on in this work, includes text generation based on retrieved knowledge, prior research including memGPT~\cite{memgpt} raft

\section{Conclusion}
{\sysname} introduces an SLO-first approach to hosting ML pipelines in which the system architecture, load levels and features such as dynamically-formed opportunistic batching all contribute to tightness of SLOs and throughput.  Central to the approach is a perspective that rather than drive latency lower and lower even at the cost of reduced throughput, we should focus on maximizing throughput so long as the SLO offered to the application meets its requirements.  As a result, our approach enables the kind of performance model shown in Figure~\ref{fig:preflmr_slo_throughput} and~\ref{fig:AudioQuery_SLO_throughput} to be offered to developers as a form of SLO contract.  Given a target SLO the runtime can be configured to elastically vary component pool sizes to manage throughput and avoid SLO misses.

Four optimizations turned out to be especially important: (1) We adopt an asynchronously pipelined architecture optimized from end-to-end to avoid copying and unnecessary locking. (2) We propose a new approach to optimistic batching that limits batch sizes based on observed behavior of the ML components in a pipeline. (3) We preprovision ML servers by loading models into GPU in anticipation of need so that when resizing does occur, a startup period of disrupted latencies and SLO misses can be avoided. (4) If available, we use RDMA networking (although our solution is shown to work well even with standard TCP). 

The combined benefits are substantial: we greatly outperform Torch Serve, but also can sustain much higher throughput for given SLOs than Ray Serve, which is today's state-of-the-art hosting product.  Interestingly, our throughput equals that of Ray Serve in situations that are not latency-limited by SLOs, pushing back on a prevailing view that prioritizing latency will invariably harm throughput.

Appendix A discusses data consistency and why it may become important in ML service deployments.  Appendix B provides additional details on our microbenchmarks for scalability of the two pipelines, and Appendix C profiles the GPU efficiencies achieved using  GRACT visualizations.
\section{Acknowledgements}
We are grateful to Microsoft, Siemens, Cisco, and NVIDIA for providing funding and resources that supported this work. We also thank Professor Stephanie Wang, Professor Christopher De Sa, Professor Mark Silberstein, and Ben Landrum for their valuable suggestions.

% In the unusual situation where you want a paper to appear in the
% references without citing it in the main text, use \nocite
\nocite{langley00}

\bibliography{citations}
\bibliographystyle{mlsys2025}

%%%%%%%%%%%%%%%%%%%%%%%%%%%%%%%%%%%%%%%%%%%%%%%%%%%%%%%%%%%%%%%%%%%%%%%%%%%%%%%
%%%%%%%%%%%%%%%%%%%%%%%%%%%%%%%%%%%%%%%%%%%%%%%%%%%%%%%%%%%%%%%%%%%%%%%%%%%%%%%
% SUPPLEMENTAL CONTENT AS APPENDIX AFTER REFERENCES
%%%%%%%%%%%%%%%%%%%%%%%%%%%%%%%%%%%%%%%%%%%%%%%%%%%%%%%%%%%%%%%%%%%%%%%%%%%%%%%
%%%%%%%%%%%%%%%%%%%%%%%%%%%%%%%%%%%%%%%%%%%%%%%%%%%%%%%%%%%%%%%%%%%%%%%%%%%%%%%
\clearpage
\appendix

\section*{A. Consistency in {\sysname}}
Consistency models are explanatory: they enable the user to reason about the scenarios that can arise during execution, which facilitates analysis of the possible behaviors that might arise in the application. We noted that  {\sysname} builds on a consistency model from the {\rdmalib} framework.   Here we briefly discuss the model in more detail and offer an example of how consistency can benefit certain ML inference and knowledge retrieval pipelines.

\begin{figure}[H]
 \centering
 \includegraphics[width=.45\textwidth]{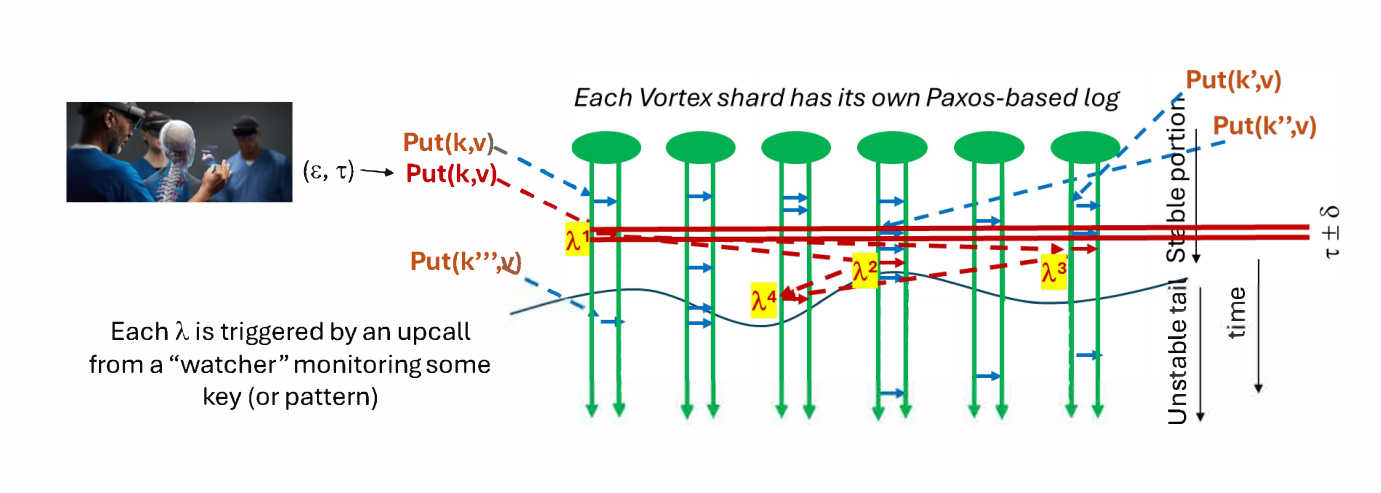}
 \caption{Medical use of AudioQuery; the yellow lambdas represent the stages of the ML pipeline.}
 \label{fig:a1}
\end{figure}

Data consistency issues arise in systems that have some form of evolving system state, which in our setting centers on updates to ML models (for example if an inference or knowledge retrieval system is dynamically learning), updates to RAG databases, and updates to other kinds of databases or data dependencies.  To make this concrete, imagine a system inspired by AudioQuery offered to physicians and supporting a continuous inflow of updates (Figure ~\ref{fig:a1}).  The updates are streams of images and other data captured by sensing devices, and  support enable audio queries and dynamic visualization of muscular-skeletal, circulatory and nerve imaging.  The visualized data could highlight possible injuries together with relevant documents (lab reports, patient history, etc).

\begin{figure}[H]
 \centering
 \includegraphics[width=.45\textwidth]{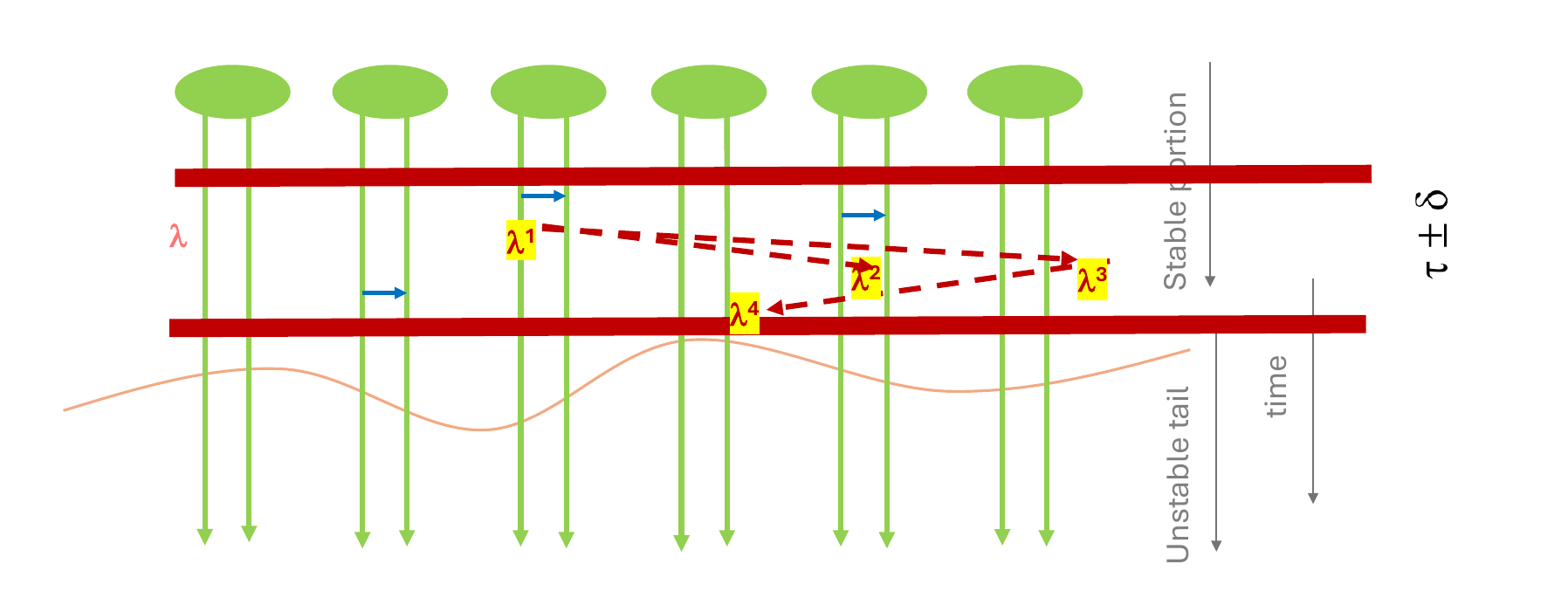}
 \caption{Zoom of a {\sysname} time-indexed {\bf get}.  Data versions are accessed along a stable consistent cut.}
 \label{fig:a2}
\end{figure}

To translate this to concrete questions about consistency in {\sysname}, we should remind ourselves that {\sysname} hosts data in the KVS, which is organized into object pools, each of which is mapped to a set of servers and then sharded under developer control.  Each shard has its own fully replicated data: as few as 2 or 3 for fault tolerance, but perhaps far more to enable elasticity.   Updates occur as {\bf put} operations, which {\sysname} routes to some member of the target shard, then issues an atomic multicast or a durable Paxos-style log append. Updates perform full-version replace, but the prior version is retained and can still be accessed:  by default {\bf get} will fetch the most current version, but time-indexing is also an option (as is iteration over a series of versions for trend analysis). Inconsistencies could include gaps in the retrieved version sequence, failure to retrieve an update that was just issued and has not yet stabilized, meaning become safely readable, or indexing errors (where the version shown for time {\bf t} is in fact from an earlier or later time).  

When we say that {\sysname} offers a formal consistency model, we are asserting that the operational description just given can also be expressed as temporal logic statements.  Safety recognizes that a {\bf put} takes time: it is issued, some period elapses during which the update is underway, and the eventually a {\em stable state} is reached in which the system starts to serve it in responses to {\bf get}.  Various additional properties then apply: that an application will read what it wrote, that data will never be read out of order or lost, and that new events can’t pop up in the stable past: once stable, events become part of a monotonic and immutable history. What if a {\bf get} were to attempt access to an unstable data item or time period?   {\sysname} would wait until the request can be safely satisfied from stable data.  The  {\rdmalib} specification is discussed in [anonymized] and analyzed using formal tools in [anonymized].   {\sysname} inherits these.

Is stabilization compatible with low latency?  This very much depends on how low the application wants latency to be, and what durability properties it seeks.  In {\rdmalib} on RDMA, experiments with 2 and 3 replication factors and smaller objects (16K) revealed stabilization delays as low as 50us, but the actual delay depends on many factors, including internal system dynamics: {\rdmalib} is aggressive about opportunistic batching, and larger writes to SSD are more efficient than a series of small ones, so when backlogs form the system shifts from an event-by-event greedy behavior to a more efficient batched behavior.  Full details can be found in ~[anonymized].

Figures~\ref{fig:a1} and \ref{fig:a2} diagram the resulting behavior as it would manifest within the {\sysname} runtime.  In Figure~\ref{fig:a1} we see {\bf put} operations saving data into a sharded object pool, as well as a {\bf trigger} that causes an upcall to $\lambda_1$ in the left-most shard.  This ML stage in turn triggers $\lambda_2$ and $\lambda_3$, and they then incast to $\lambda_4$.  Implied but not shown are the time-indexed {\bf get} operations issued by those lambdas, but the intent is that all four ML stages use data along a deterministic, consistent ``snapshot’’ across the system: the same requests will always return the same results.  

Figure~\ref{fig:a1} visualizes all of this:  we see that {\rdmalib} imposes a barrier (curved line) so that {\sysname} {\bf get} requests will not glimpse the potentially incomplete transitional states associated with {\bf put} operations still underway.  The curved line represents the stability threshold: older {\bf put} operations have stabilized, where as newer ones are still pending.  The guarantee extends to timestamps too: once stable, {\rdmalib} will never insert a {\bf put} with an older timestamp into the stable portion of the log (instead, it rejects such a {\bf put} as ``too old’’). 

Of course, any distributed system must grapple with limits on clock synchronization, as emphasized in Figure~\ref{fig:a2}.  Here we zoom in and see that when a {\bf get} accesses data at time {\bf t}, there might still be multiple candidate object versions that could correspond to {\bf t}.  Any time must be interpreted plus or minus the possible clock synchronization skew.   The {\rdmalib} policy is to return the most current stable version that is not later than the requested time.   For reads of different objects that all fall within a temporal window around {\bf t}, {\rdmalib} employs a representation that combines clock time with a form of Lamport-style causality tracking,  using an algorithm described in~\cite{FFFS}.  The effect is that {\bf get} operations occur along a stable consistent cut~\cite{consistentcuts}.

What if the application needs to update multiple objects and desires that this be atomic?  If the objects reside on the same shard, this is trivial: {\bf put} allows a list of KV pairs to be specified, and just performs the updates in an uninterruptible, atomic, manner.  But in fact there is a fairly simple application-layer strategy for extending this behavior into full transactions across multiple shards.  

The basic idea here is old, but this form of it is similar to an algorithm used in the Heron system~\cite{Heron} that in turn builds on Chain Replication~\cite{chain}:  The application first must pre-execute the transaction (without any form of locking), building a speculative list of objects to be read or updated.  Then in a second stage, it traverses shards that were accessed (for reads as well as for speculated writes) in order, left-most shard to right-most shard.  As a transaction visits a shard, it temporarily locks any objects it will access at that shard, and logs the read and update set for objects at that shard to a write-ahead log.  When the transaction reaches the tail of the chain, it will have confirmed (1) that nothing it depended upon changed since the speculative execution, and (2) the updates are safe to commit.  At the tail, we log a record that the transaction committed, then perform the commit, rightmost shard to leftmost shard, unlocking the accessed objects.  Conversely, if it discovers that some other transaction updated some of the objects it wishes to read or write, the transaction aborts, shard by shard from right to left.  Finally, if a transaction encounters a locked object, it waits for the prior transaction to finalize.  Recovery from failure is straightforward.

We do not currently offer this as a feature in {\sysname}, but it would be trivial to do so if future users require it.
Thus, the  {\sysname} KVS can support a very powerful form of database, although we currently use it in a  simplified way.  Moreover, by reading only from stable data in an immutable state, we obtain determinism and the assurance that a transaction that reads multiple objects, even from separate shards, will see a consistent cut.  This  matches closely with a database model referred to as serializable snapshot isolation~\cite{SerializableSnapshotIsolation}.

Revisiting our medical scenario,  the ML physician's assistant will never base a response or action on a transiently unstable state with missing data, causal ordering gaps, ``mashups’’ that combine data from two transactions, etc.  Clearly this is a desirable guarantee.  Why, then, did we characterize consistency as a debatable requirement, in Section~\ref{sec:platform_design}?  The real issue centers in part on how often the mix of real-time updates, ML with genuine safety needs, and a need to respond under intense time-pressure will arise, and in part on the degree of developer and consumer awareness of the roles that data inconsistencies can play in ML malfunctions.  

In our own prior work, we encountered and described such a puzzle arising in an experimental R\&D project we undertook for a consortium of operators of the US smart bulk (large-scale) power grid.  Power grids are heavily regulated and our work showed that while cloud hosting of power grid systems is possible, data inconsistencies arising in standard cloud file systems could prevent proper behavior, and we offered a user-level storage solution as a work-around [anonymized].  The community embraced the research, yet regulatory policies were not revised to express an obligation that power control systems operate on provably consistent input data.  But this doomed our work: Power grids are so heavily regulated that dependency on properties that are not obligatory is often questioned as a safety risk.  

Similarly, while the medical scenario we used would be subject to regulatory scrutiny and hence required to anticipate and protect against such threats, it is entirely possible that regulators could be be satisfied by experimental demonstrations of safety, perhaps accompanied by a discussion of heuristic mechanisms, such as a self-check to sense and retry requests that seem to have encountered a data instability.  We have seen this same situation in other regulated industries, such as as ML for air traffic control or fly-by-wire aircraft.   Self-driving cars deploy ML in safety-critical settings and accidents have occurred, but none has been attributed to data inconsistencies that caused dangerous mistakes.

\captionsetup[subfigure]{justification=centering, singlelinecheck=false}
\begin{figure*}[!htbp]
    \centering
    % Row 1 — Pipeline1 (4 plots)
    \begin{subfigure}[b]{0.24\textwidth}
        \centering
        \includegraphics[width=\textwidth]{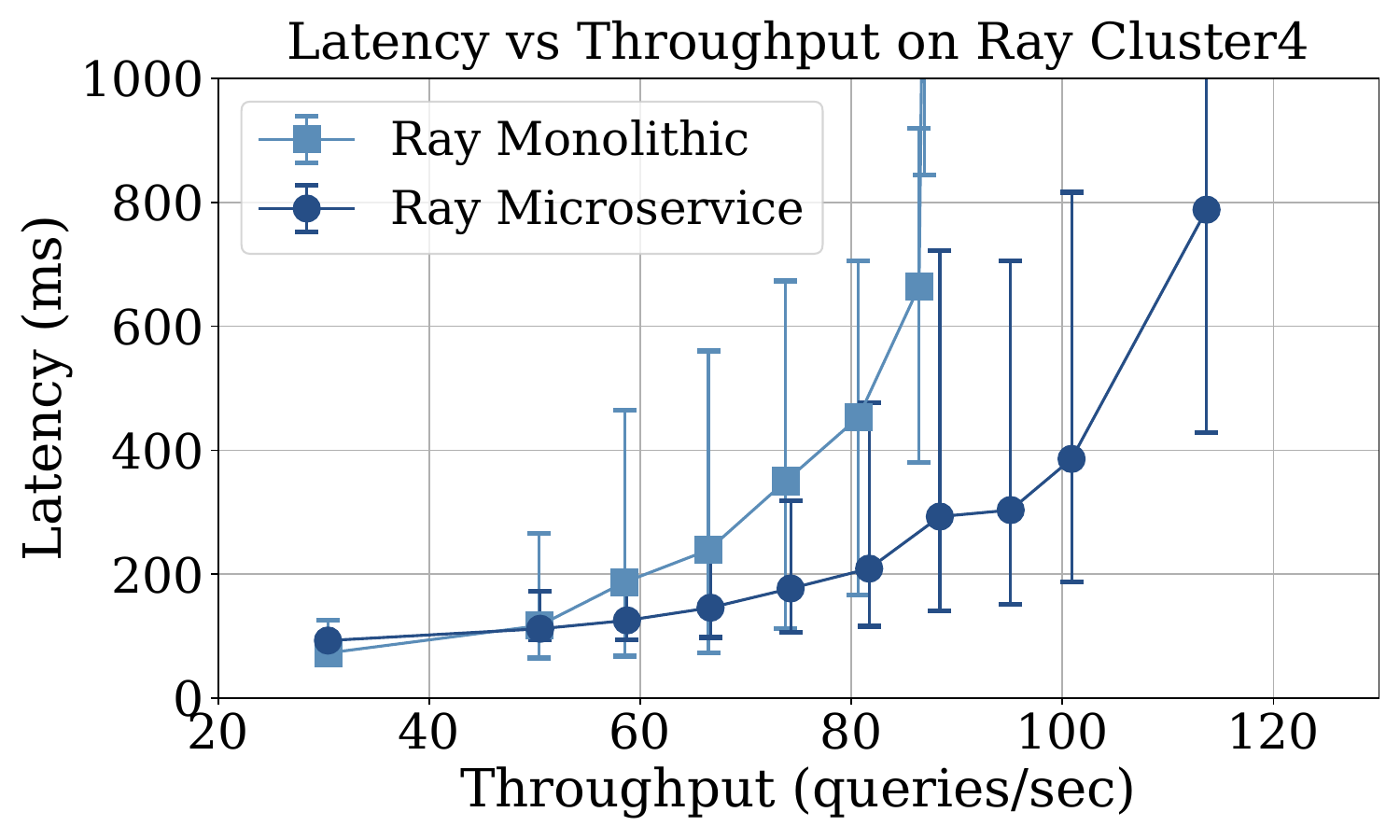}
        \caption{PreFLMR, \\ 4 nodes \textbf{Ray Serve}}
        \label{fig:ppl1_ray_mono_vs_micro_cluster4}
    \end{subfigure}
    \begin{subfigure}[b]{0.24\textwidth}
        \centering
        \includegraphics[width=\textwidth]{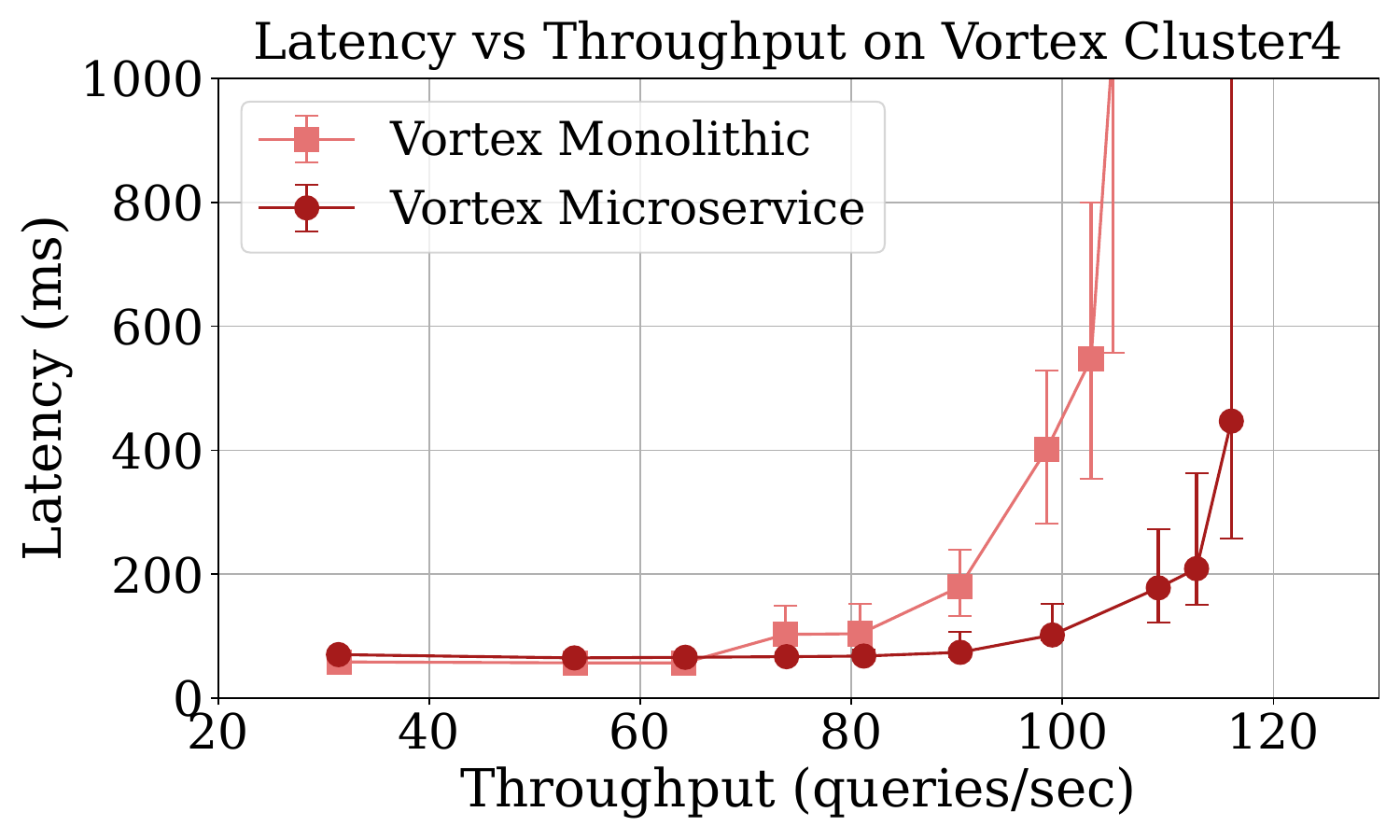}
        \caption{\centering PreFLMR, \\ 4 nodes \textbf{{\sysname} (RDMA)}}
        \label{fig:ppl1_vortex_mono_vs_micro_cluster4}
    \end{subfigure}
    \begin{subfigure}[b]{0.24\textwidth}
        \centering
        \includegraphics[width=\textwidth]{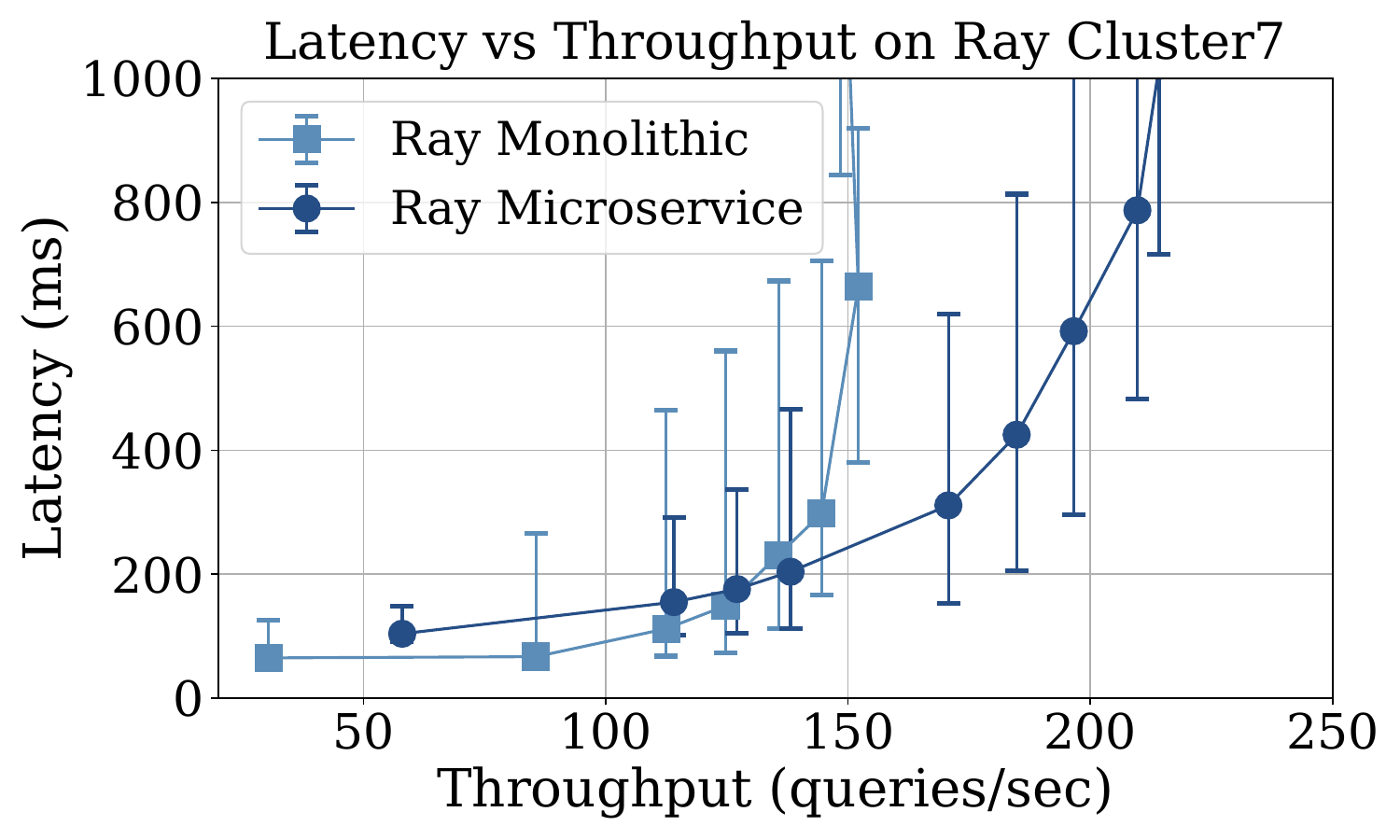}
        \caption{\centering PreFLMR, \\ 7 nodes \textbf{Ray Serve}}
        \label{fig:ppl1_ray_mono_vs_micro_cluster7}
    \end{subfigure}
    \begin{subfigure}[b]{0.24\textwidth}
        \centering
        \includegraphics[width=\textwidth]{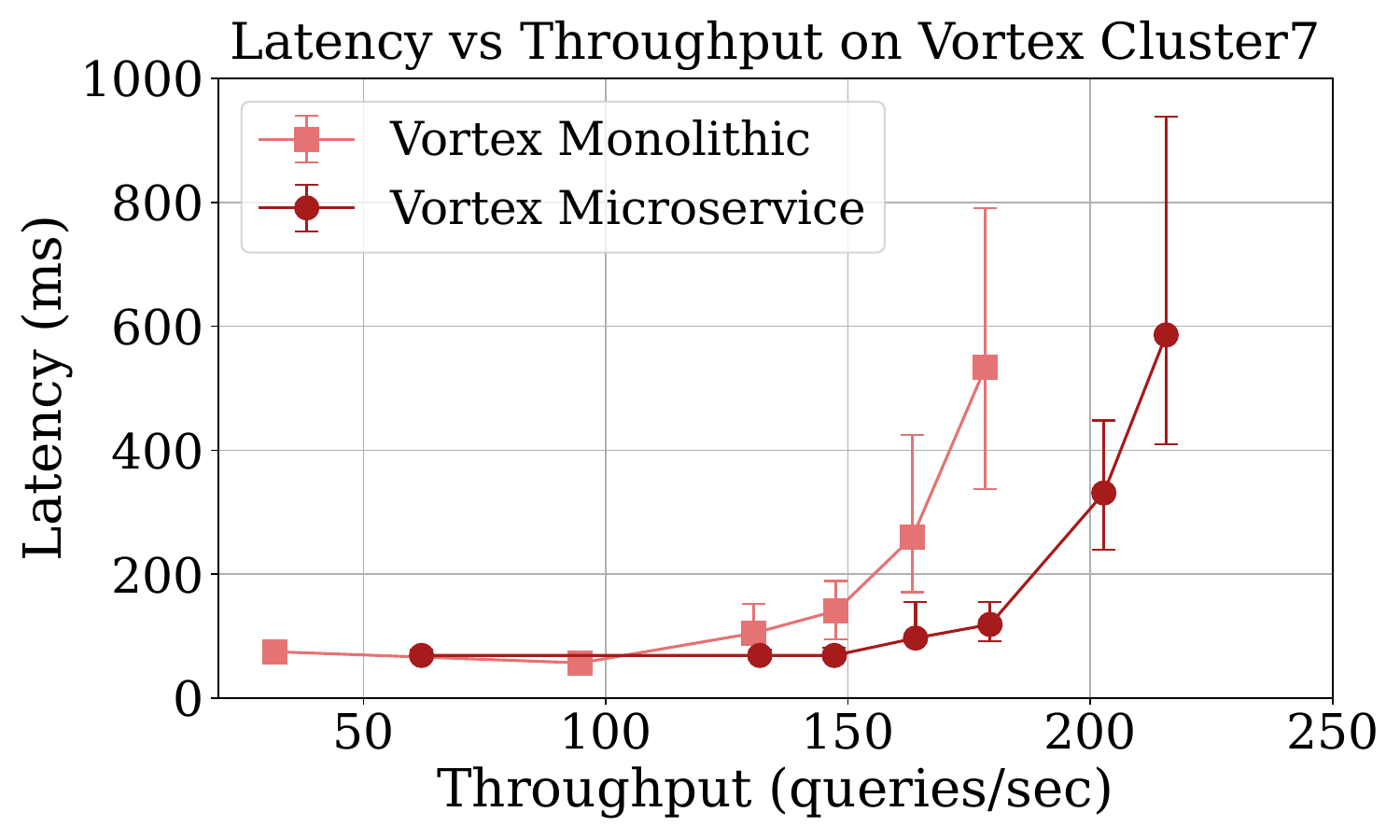}
        \caption{\centering PreFLMR, \\ 7 nodes \textbf{{\sysname} (RDMA)}}
        \label{fig:ppl1_vortex_mono_vs_micro_cluster7}
    \end{subfigure}

    \vspace{0.5em} % Add vertical space between rows

    % Row 2 — Pipeline2 (5 plots)
    \begin{subfigure}[b]{0.24\textwidth}
        \centering
         \includegraphics[width=\textwidth]{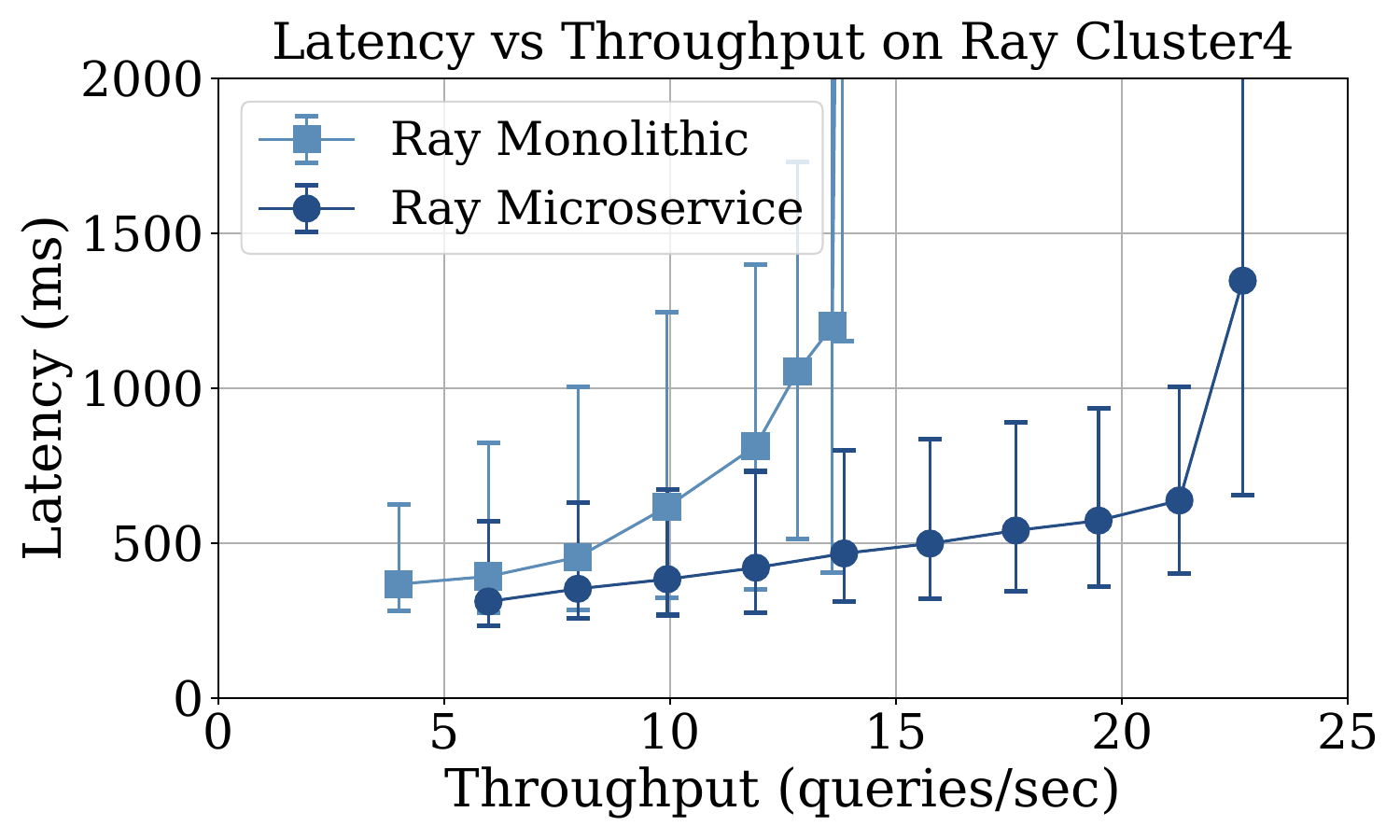}
         \caption{AudioQuery, \\ 4 nodes \textbf{Ray Serve}}
         \label{fig:ppl2_ray_mono_vs_micro_cluster4}
     \end{subfigure}
     \begin{subfigure}[b]{0.24\textwidth}
         \centering
         \includegraphics[width=\textwidth]{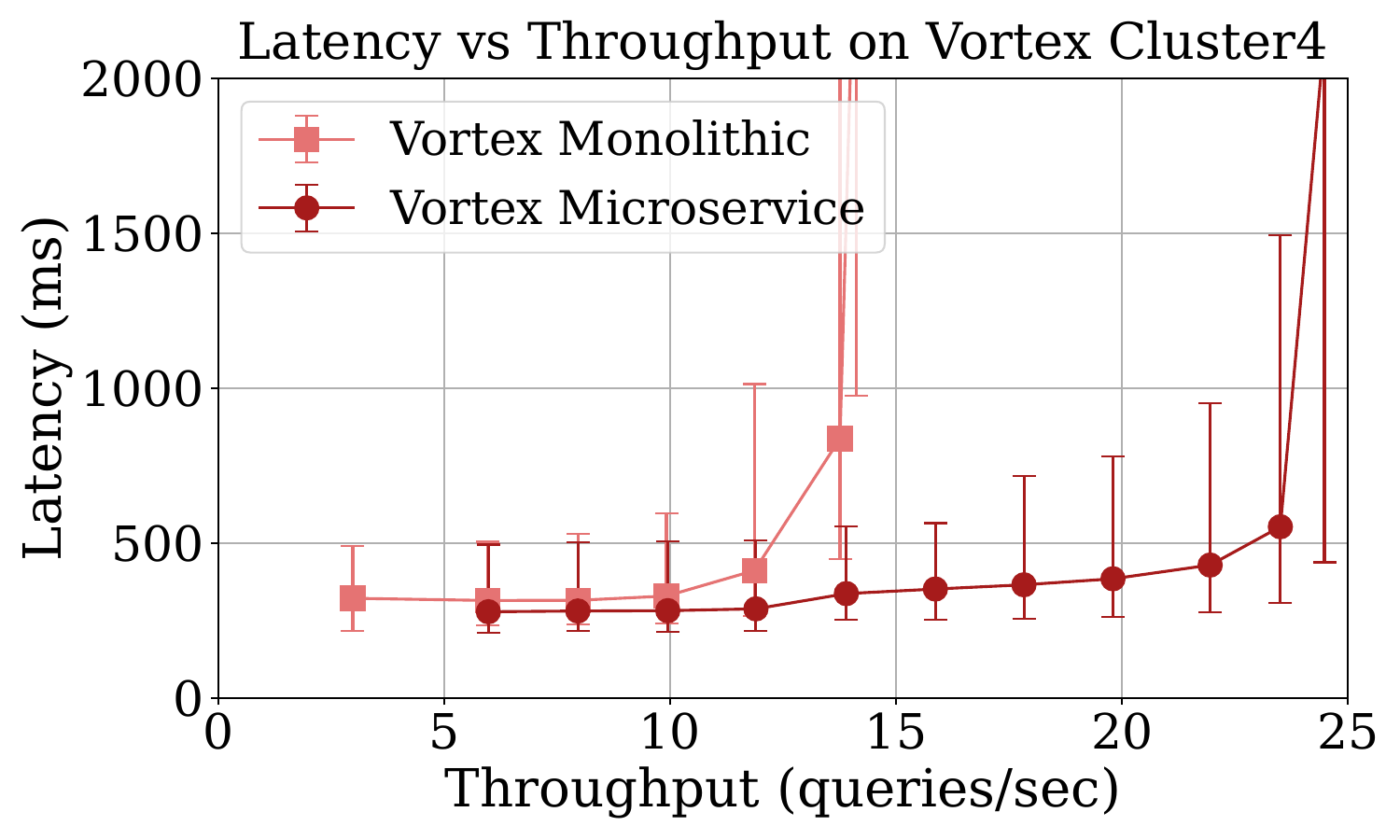}
         \caption{AudioQuery, \\ 4 nodes \textbf{{\sysname} (RDMA)}}
         \label{fig:ppl2_vortex_mono_vs_micro_cluster4}
     \end{subfigure}
     \begin{subfigure}[b]{0.24\textwidth}
         \centering
         \includegraphics[width=\textwidth]{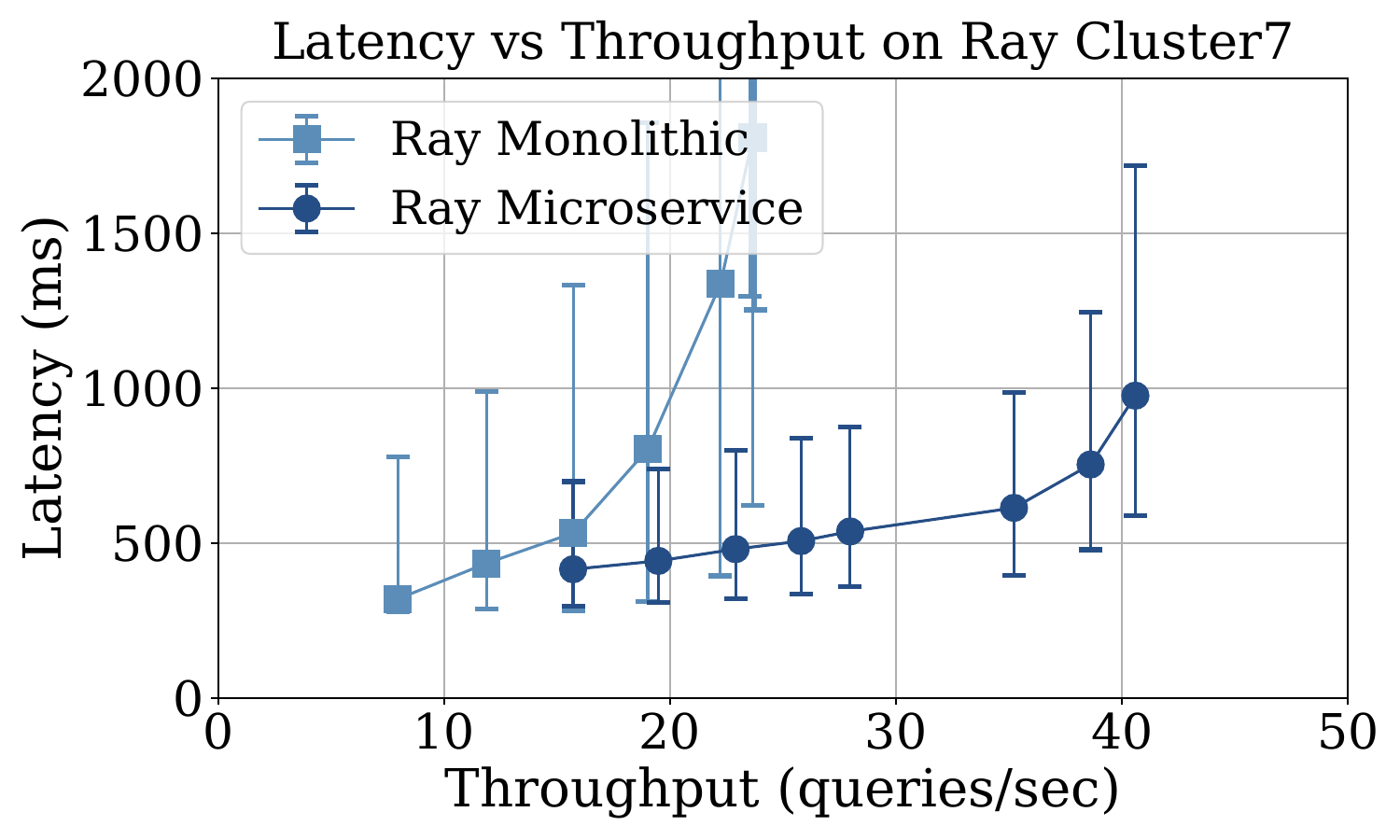}
         \caption{AudioQuery, \\ 7 nodes \textbf{Ray Serve}}
         \label{fig:ppl2_ray_mono_vs_micro_cluster7}
     \end{subfigure}
     \begin{subfigure}[b]{0.24\textwidth}
         \centering
         \includegraphics[width=\textwidth]{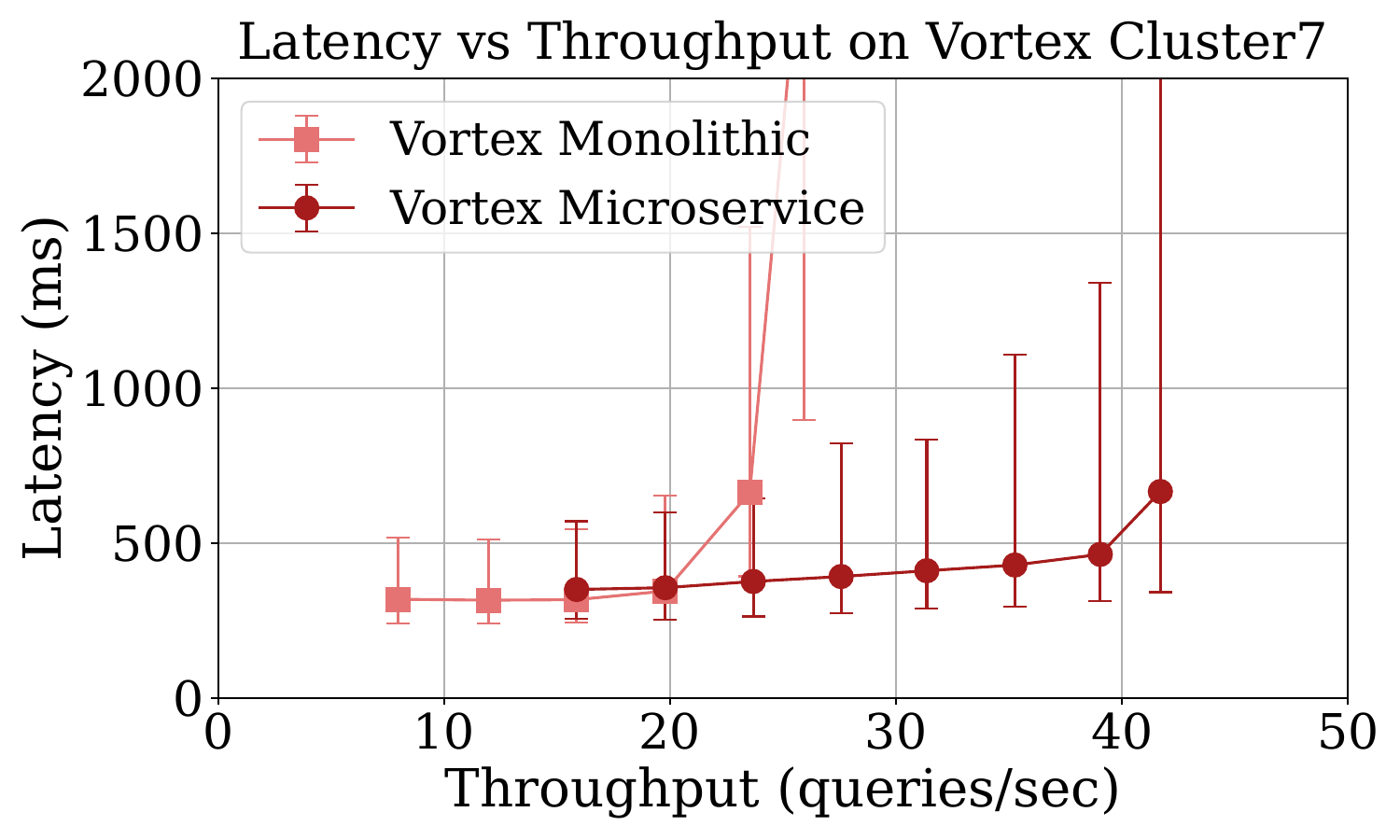}
         \caption{ AudioQuery, \\ 7 nodes \textbf{{\sysname} (RDMA)}}
         \label{fig:ppl2_vortex_mono_vs_micro_cluster7}
     \end{subfigure}

    \vspace{.15in}\caption{Latency (showing median, 5\% and 95\% latency range) as we vary the input query rate and cluster size.}
    \label{fig:mono_vs_micro_deployment}
\end{figure*}

We view this as a chicken-and-egg problem.  Most existing ML platforms ignore storage layer consistency, or run monolithically on a single server (individually hosted file systems guarantee ``read what I wrote'' consistency).  Cloud storage systems do exhibit inconsistency under update-reread timing pressure, but there is usually a long delay between when an update is done and when data is reread, hence ample time for caches to regain coherency.  Cloud vendors often offer consistent data-storage products, but they are widely viewed as slow and the market is limited.  

Thus even today's safety-critical ML solutions tend not to consider infrastructure-induced data inconsistency to be a priority.  While {\sysname} does possess stronger guarantees (and they have no real performance costs), application-level demand for framework consistency guarantees will probably remain limited until (or unless) experimental studies of ML hallucinations and errors begin to highlight inconsistency as a significant root cause responsible for a non-trivial rate of serious errors.  The situation is reminiscent of an early era in computing systems that deployed a generation of functionally impressive yet buggy solutions.  

That era only  ended when a series of research papers and systems were published on topics like how often and why systems fail, the pernicious role of concurrency bugs, better methodologies for writing and verifying correct concurrent programs, and sophisticated code analysis tools and formal methods to improve system quality.  With growing scrutiny and awareness, consumer demand began to shift, and today's databases and operating systems are far more robust.  We routinely talk about models such as atomicity, linearizability and serializability, and embed easily-used tools and libraries that embody these models into software-development environments.  Presumably, the ML deployment community will eventually reach that same point.  The question is when it will happen.  Our conjecture is that when ML as a service is widely deployed and some important instances experience high update rates, awareness of the issue will grow and the kinds of properties {\sysname} inherits from {\rdmalib} will be more widely recognized as valuable.

\section*{B. Scaling Microservice vs. Monolithic Deployment to More Servers}

Figure~\ref{fig:mono_vs_micro_deployment} offers more detail on how the  ML services we considered scale in various configurations.  We run PreFLMR and AudioQuery on Ray Serve and {\sysname}, comparing the best configurations for each in microservice and monolithic deployments.  Note that whereas Ray Serve only supports TCP deployment, the best configuration of {\sysname} is its RDMA configuration.  We did collect a data set that includes all of these choices (it was used to create Figure~\ref{fig:latency_vs_throughput_all_configurations}, which includes {\sysname} on TCP), but because our work argues that SLO-oriented services should favor RDMA when possible, we omit the {\sysname} TCP data here.  For these pipelines RDMA accounts for about half of our advantage over Ray Serve, as was seen in Figure~\ref{fig:latency_vs_throughput_all_configurations}. 

The microservice deployment consistently outperforms the monolithic setup, achieving throughput improvements of \textbf{1.31$\times$} and \textbf{1.56$\times$} for Ray, and \textbf{1.12$\times$} and \textbf{1.10$\times$} for {\sysname} on the PreFLMR and AudioQuery pipelines, respectively. These results demonstrate that microservice architectures enable higher throughput across diverse workloads.

% These curves thus show low-rate behavior, when queries can be handled more or less one by one, as well as much higher rates where batching is critical.  We vary the cluster size to study a modest form of scalability (reservations of larger numbers of nodes was impractical on CloudLab).  For each experiment, we increase the presented load until congestion-control mechanisms prevent further load increase. As shown in Figure~\ref{fig:mono_vs_micro_deployment}, we obtain two sets of eight plots (one set per pipeline). Each plot shows the distribution of measured latency as a function of system throughput. 

As shown in the graphs, monolithic deployments do achieve the lowest end-to-end latency when throughput is extremely low. This is because all pipeline components run on the same machine within the PyTorch address space, minimizing data transfer overhead. While Python may introduce some data copying, PyTorch is generally efficient in avoiding unnecessary GPU memory movement when the next operation is also GPU-bound.  The issue is that ML services would rarely see just one or two queries at a time.  At even slightly higher loads, latency degrades sharply when compared with the microservice deployments.  The microservice architectures also benefit from packing, and if multiple
workflows can share a component pool, would see further efficiency benefits.  Thus, while a monolithic deployment would give the very lowest latency, it does so at higher cost.

For the PreFLMR pipeline on a 4-node cluster, Ray Serve’s microservice deployment achieves a median latency approximately \textbf{1/3} that of its monolithic counterpart (Figure~\ref{fig:ppl1_ray_mono_vs_micro_cluster4}), while {\sysname} achieves a latency reduction of \textbf{1/2} to \textbf{1/4} (Figure~\ref{fig:ppl1_vortex_mono_vs_micro_cluster4}). This trend holds as the cluster scales from 4 to 7 nodes (Figures~\ref{fig:ppl1_ray_mono_vs_micro_cluster7} and~\ref{fig:ppl1_vortex_mono_vs_micro_cluster7}).
% For PreFLMR pipeline running on Ray, the ratio of median latency between microservice and monolithic is \textbf{1/2} on 7 nodes setting, same for {\sysname}. 
% Workflow also has affect on the improvement. 
The benefits of the microservice architecture are more pronounced for AudioQuery (Figure~\ref{fig:ppl2_ray_mono_vs_micro_cluster4}-~\ref{fig:ppl2_vortex_mono_vs_micro_cluster7}) than PreFLMR (Figure~\ref{fig:ppl1_ray_mono_vs_micro_cluster4}-\ref{fig:ppl1_vortex_mono_vs_micro_cluster7}). This is due to the larger intermediate data sizes in PreFLMR compared to AudioQuery, which mostly passes text fragments from stage to stage.  As a result, microservice deployments not only achieve up to \textbf{2$\times$} the throughput of monolithic setups but also maintain low latency even at high query rates.

\section*{C. GPU Utilization Efficiency}
In this appendix we profile the GPU resource utilization of microservice and monolithic deployments of PreFLMR.  The setup is the one used in Section~\ref{sec:challenges}, Figures~\ref{fig:mono_vs_micro_timeslines} and \ref{fig:mono_vs_micro_config}. 
Our evaluation employs a GRACT metric: a measure of the fraction of time the GPU compute units were actively performing computational tasks~\cite{MIGPerf,profile_training_euromlsys}.

\begin{figure}[ht]
    \centering
    \includegraphics[width=0.4\textwidth]{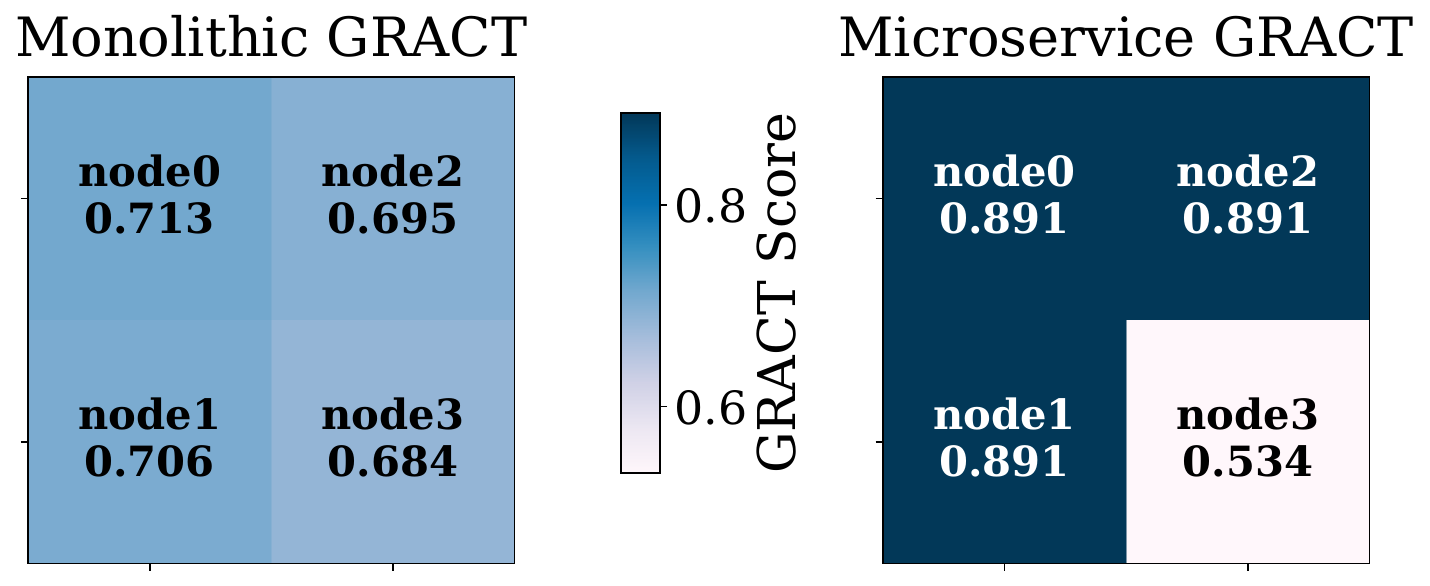}
    \caption{Average percent of graphics or compute resource active (GRACT) 
    for PreFLMR on 4 nodes running at the highest sustainable load on \textbf{{\sysname}}.  Darker is better.}
    \label{fig:gract_heatmap}
\end{figure}

Figure~\ref{fig:gract_heatmap} presents the GRACT visualization for both microservice and monolithic deployments running at their respective peak throughputs on {\sysname}. 
In the monolithic setup each node runs the entire pipeline.
As a result, all components (not just the most GPU intensive ones) remain loaded in GPU memory throughout the execution.  Our test includes the start of the query stream: a period when node 0 begins work first, then node 1, 2, and finally node 3.  We saw in Figure~\ref{fig:mono_vs_micro_timeslines} that monolithic deployment hinders efficiency, and the GRACT bears out that intuition (left side).  In contrast, the microservice deployment achieves not just better packing efficiency, but also improved computational efficiency: the three nodes running the vision encoder task are highly utilized, while the fourth node (which runs less GPU-intensive components) is less heavily utilized.

%%%%%%%%%%%%%%%%%%%%%%%%%%%%%%%%%%%%%%%%%%%%%%%%%%%%%%%%%%%%%%%%%%%%%%%%%%%%%%%
%%%%%%%%%%%%%%%%%%%%%%%%%%%%%%%%%%%%%%%%%%%%%%%%%%%%%%%%%%%%%%%%%%%%%%%%%%%%%%%

\end{document}